\documentclass[pdflatex,sn-mathphys-num]{sn-jnl}

\usepackage{graphicx}%
\usepackage{multirow}%
\usepackage{amsmath,amssymb,amsfonts}%
\usepackage{amsthm}%
\usepackage{mathrsfs}%
\usepackage[title]{appendix}%
\usepackage{xcolor}%
\usepackage{textcomp}%
\usepackage{manyfoot}%
\usepackage{booktabs}\let\cline\cmidrule
\usepackage{algorithm}%
\usepackage{algorithmicx}%
\usepackage{algpseudocode}%
\usepackage{listings}%
\usepackage{tabularx} 
\usepackage{graphicx}
\usepackage{multirow}
\usepackage{threeparttable}
\usepackage[caption=false]{subfig}
\usepackage{adjustbox}
\usepackage{arydshln}
\usepackage{todonotes}

\begin{document}

\title[Urban Delivery Performance]{Urban context and delivery performance: Modelling service time for cargo bikes and vans across diverse urban environments}

\author[1]{\fnm{Maxwell} \sur{Schrader}}\email{mcschrader@crimson.ua.edu}
\equalcont{These authors contributed equally to this work.}

\author[2]{\fnm{Navish} \sur{Kumar}}\email{navish.kumar@unibas.ch}
\equalcont{These authors contributed equally to this work.}

\author[3]{\fnm{Esben} \sur{Sørig}}\email{esben@kale.ai}

\author[3]{\fnm{Soonmyeong} \sur{Yoon}}\email{chris@kale.ai}

\author[4]{\fnm{Akash} \sur{Srivastava}}\email{akashsri@mit.edu}

\author[4]{\fnm{Kai} \sur{Xu}}\email{xuk@mit.edu}

\author[5]{\fnm{Maria} \sur{Astefanoaei}}\email{msia@itu.dk}

\author*[3]{\fnm{Nicolas} \sur{Collignon}}\email{nicolas@kale.ai}
\equalcont{These authors contributed equally to this work.}

\affil[1]{\orgdiv{Department of Mechanical Engineering}, \orgname{University of Alabama}, \orgaddress{\country{USA}}}

\affil[2]{\orgdiv{Department of Mathematics and Computer Science}, \orgname{University of Basel}, \orgaddress{ \country{Switzerland}}}

\affil*[3]{\orgname{Kale AI}, \orgaddress{ \city{London}, \country{UK}}}

\affil[4]{\orgdiv{MIT-IBM Watson AI Lab},\orgaddress{\street{Cambridge}, \state{Massachusetts}, \country{USA}}}

\affil[5]{\orgdiv{Data-intensive Systems and Applications}, \orgname{IT University of Copenhagen}, \orgaddress{ \country{Denmark}}}




\abstract{
    \textbf{Purpose:} Light goods vehicles (LGV) used extensively in the last mile of delivery are one of the leading polluters in cities. Cargo-bike logistics and Light Electric Vehicles (LEVs) have been put forward as a high impact candidate for replacing LGVs. Studies have estimated over half of urban van deliveries being replaceable by cargo-bikes, due to their faster speeds, shorter parking times and more efficient routes across cities. However, the logistics sector suffers from a lack of publicly available data, particularly pertaining to cargo-bike deliveries, thus limiting the understanding of their potential benefits. Specifically, service time (which includes cruising for parking, and walking to destination) is a major, but often overlooked component of delivery time modelling. The aim of this study is to establish a framework for measuring the performance of delivery vehicles, with an initial focus on modelling service times of vans and cargo-bikes across diverse urban environments.
    
    \textbf{Methods:} 
    We introduce two datasets that allow for in-depth analysis and modelling of service times of cargo bikes and use existing datasets to reason about differences in delivery performance across vehicle types. We introduce a modelling framework to predict the service times of deliveries based on urban context. 
    We employ Uber’s H3 index to divide cities into hexagonal cells and aggregate OpenStreetMap tags for each cell, providing a detailed assessment of urban context. Leveraging this spatial grid, we use GeoVex to represent micro-regions as points in a continuous vector space, which then serve as input for predicting vehicle service times. 
    We show that geospatial embeddings can effectively capture urban contexts and facilitate generalizations to new contexts and cities. Our methodology addresses the challenge of limited comparative data available for different vehicle types within the same urban settings. 

    \textbf{Results:} Our findings indicate a significant impact of urban context on the performance of vans, with cargo-bikes showing less sensitivity to these variations. While there is strong suggestive evidence for performance benefits of cargo-bikes in dense urban areas, our analysis is constrained by the lack of corresponding data for vans and cargo-bikes operating under similar urban conditions and logistics scenarios. The preliminary results of our models demonstrate some capability to generalize to new and unseen urban contexts, suggesting a promising path for future research.

}




\maketitle

\section{Introduction}\label{sec1}

Today, transport accounts for around 30\% of a city's carbon emissions, and is expected to grow at a faster rate than any other sector in the coming decade~\cite{creutzig2015transport}. One of the drivers behind this acceleration is the increasing demand for faster logistics due to the boom of the e-commerce industry. 
By 2025, the number of packages delivered around the world is expected to climb to 200 billion, up from fewer than 90 billion in 2018~\cite{statista21}.
Given the current trends, it is estimated that both the amount of CO2 emissions and traffic congestion caused by urban last-mile deliveries will increase by 30\% by 2030 in the top 100 cities globally (and an additional 6 million tonnes of CO2 will be emitted compared to 2019)~\cite{deloison2020future}.

The single most challenging aspect for logistics operators in meeting this exceptional rise in demand comes down to the last mile of the delivery~\cite{savelsbergh201650th}, with some reporting it accounts between 40 to 60\% of the total delivery cost of a parcel~\cite{bachofner2022city, Statista2023}. Recently, cargo bike logistics has emerged as a competitive and environmentally sustainable solution for replacing LGV fleets in cities, promising improved efficiency in dense urban areas~\cite{gruber2014new, sheth2019measuring,van2018city,wrighton2016cyclelogistics,possible21}.

A recent study found that 67\% of daily van operations of a large logistics operator in Paris could be substituted by cargo-bikes at no extra cost~\cite{robichet2022first}. This study introduced a comprehensive framework to analyze the trade-off between the operational costs of vans and cargo-bikes, and the cost of micro-hub infrastructure~\footnote{A micro-hub is a small storage facility in an urban area that receives deliveries for local distribution, often through sustainable modes like cargo-bikes.}. Their model simplifies vehicle performance to a single constant (delivery per hour), and only used weight as an upper capacity limit due to a lack of volume data.

While the results strikingly demonstrate the large potential of cargo-bikes, they also represent a notable limitation commonly found in cargo-bike logistics research. Specifically, there exists a wide variation in performance assumptions regarding average vehicle speeds and cargo-bike load capacities. For example, Lee et al.\cite{lee2019courier} assume a cargo-bike capacity of 350 kg and an average speed of 10 km/h (while admitting this is not based on any data), compared to a speed of 30 km/h for an LGV. Conversely, Elbert et al.\cite{elbert2020urban} ignore the speed of the vehicle, deciding instead to focus on service time, with 2 min for the cargo-bike vs up to 8 min for the LGV, while assuming a cargo-bike capacity of 125 kg. In Arnold et al.~\cite{arnold2018simulation}, a cargo-bike moves at 12 km/h vs 17 km/h for an LGV, but no differentiation is made for service time. This variation in assumptions can be attributed to the scarcity of reliable data and the complexities of modeling fine-grained delivery scenarios, which ultimately points to a significant gap in our understanding. 

In this paper, we specifically focus on service time, given its significant impact on overall delivery efficiency, especially in dense urban areas. Service time, also referred to as stop time, dwell time, delivery time in the literature, typically refers to the time needed to cruise for parking, unload parcels, walk between the delivery vehicle and the destination address, and time handing off the delivery. We decide to study service time as it is an often disregarded component of route optimisation and a major performance indicator, that is also heavily dependent on the type of vehicle used. For example, a van may be delayed by traffic congestion before potentially needing extended cruising for available parking, in turn leading to a longer distance on foot to the delivery point. This can represent a significant part of a delivery driver's day. In contrast, a cargo-bike, due to its size and maneuverability, can move through traffic more effectively and often finds parking closer to the final delivery location.

The importance of accurate, data-driven decision-making is paramount in an industry known for high pressure and tight margins. Yet, data collection processes vary widely in maturity and many incumbent companies grapple with data quality issues across the logistics chain, which hampers their data science efforts. These challenges in data collection and interpretation within the logistics industry extend beyond long-term strategic decisions, such as fleet composition, and affect daily operational issues too. Operators still lack comprehensive tools that accurately capture the intricate facets of driver behaviour and the variable efficiency of vehicles in real-world scenarios~\cite{almrcc}. Recognizing the need for a more fine-grained understanding of delivery operations, we introduce datasets as well as a framework for modeling delivery efficiency that takes into account vehicle type and urban micro-regions. The main contributions of the paper are as follows:

\begin{itemize}
\item We present novel datasets that capture detailed service time information for both cargo-bikes and vans across diverse urban environments. These datasets provide a first step towards a better understanding of delivery efficiency in urban areas.
\item Through rigorous analysis of these datasets, we highlight the contrasting operational profiles of cargo-bikes and vans, emphasizing the necessity for vehicle-specific modeling. Furthermore, we demonstrate the significant impact of urban context on service times.
\item To effectively capture the influence of urban context, we propose leveraging embedding models to generate vector representations of urban micro-regions. We showcase the relevance and utility of these representations in the context of predicting service times for last-mile deliveries.
\item Building upon these insights, we develop vehicle-specific service time prediction models that incorporate the vector representations of urban context. These models enable more accurate and context-aware predictions of delivery efficiency.
\end{itemize}

\section{Related works}\label{related-works}

Studying the behaviour of delivery vehicles and their efficiency across different urban areas is a multi-faceted problem, that is still not well understood. A number of studies have estimated delivery vehicle performance based on either aggregated values, such as average speed, capacity, service time \cite{lee2019courier,elbert2020urban,arnold2018simulation}, or on GPS data showing the full routes and timestamps of deliveries \cite{greaves2008collecting, ma2011processing, zhao2011evaluating, yang2014urban}. The later works focused on truck deliveries and analysed performance measures such as speed, round duration, distance traveled, number of stop locations. 

With the rise of e-commerce and the growing awareness of the negative impact of vans on cities, there is rising interest in understanding the potential of integrating cargo bikes in commercial deliveries in terms of efficiency \cite{arnold2018simulation, ritzer2023traffic, sheth2019measuring, llorca2021assesment, perboli2019parcel, elbert2020urban, zhang2018simulation}. However, only few of these have considered service time as a factor in assessing efficiency. 
For example, \citeauthor{arnold2018simulation} \cite{arnold2018simulation} chooses a fixed value for service time, estimated from a dataset of van deliveries, while only \citeauthor{possible21}  \cite{possible21}, \citeauthor{elbert2020urban} \cite{elbert2020urban} and \citeauthor{perboli2019parcel} \cite{perboli2019parcel} choose differentiating services times between vans and cargo bikes. 

One reason why service time has been largely ignored in vehicle performance comparison studies is the difficulty to accurately identify stop locations from GPS traces. This is due to a lack of direct correspondence between stop locations and delivery addresses \cite{yang2014urban, ma2011processing} (for example one stop can correspond to multiple deliveries). \citeauthor{conway2017cargo} \cite{conway2017cargo} used spot speeds collected from GPS-equipped cargo cycles and trucks to estimate corridor speed, stopped-time delay to travel time ratio, and parking time. Their study revealed that the cargo cycles had a lower stopped-time delay than motorised vehicles, suggesting they are more reliable, particularly in congested conditions \cite{conway2017cargo}. However the data was collected from only 2-3 vehicles for each type and in a limited geographical area, making it difficult to argue about the generalisability of the results.

Another challenge in estimating service time is predicting the duration of cruising for parking. A main issue in estimating delivery times in traditional van last mile logistics is that parking outside the delivery address is often impossible. Reasons for this include parking restrictions, availability, and traffic congestion \cite{nguye2019optimising}. 
In their review, \citeauthor{ghizzawi2024modelling} \cite{ghizzawi2024modelling} emphasise the complexity and importance of parking choices in urban logistics, including the ramifications of illegal parking. This behaviour not only imposes substantial costs on operations, evidenced by millions in parking fines, but also affects city life, contributing to traffic congestion and jeopardizing the safety of other road users, including cyclists.

In terms of costs on operations due to cruising for parking, studies found that delivery drivers spend on average 5.8 minutes and 24 minutes searching for each parking spot in Seattle and New York City, respectively ~\cite{dalla2020commercial, holguin2016impacts}. In their Seattle study, \citeauthor{dalla2020commercial}~\cite{dalla2020commercial} found that on average, cruising for parking accounted for 28\% of the total trip time between parking locations. Moreover, due to long distances between parking and delivery location, walking to destination becomes a significant factor in the total delivery time. A 2018 study in London highlighted that walking can account for 62\% of the total van round time~\cite{london_ftc2050} due to limited suitable kerbside parking space in urban centres. 

\citeauthor{ghizzawi2024modelling} \cite{ghizzawi2024modelling} conducted a scoping review of research on modelling parking behavior of commercial vehicles (specifically, trucks and vans), and highlight the need for more research and data collection. Overall, they point at four categories influencing parking behaviour (which we correlate to service time performance):

\begin{itemize}
\item Decision-maker, vehicle and trip attributes (e.g. experience of the delivery staff, vehicle size, time-of-day, industry sector)
\item Parking location attributes (e.g. parking type, cost, capacity)
\item Activity and shipment attributes (e.g. delivery size, commodity type)
\item Built environment attributes (e.g. population density, employment, number and type of establishments)
\end{itemize}

These findings collectively highlight parking as a critical factor in the operational efficiency of commercial vehicles in city environments. 

Despite this, optimisation tools largely ignore the search time for parking in route planning~\cite{reed2021does}. \citeauthor{reed2021does} show that by including parking time in the objective for route optimization drastically changes the routes chosen and the completion time of the deliveries. However, choosing when and where to park remains a decision made by the driver~\cite{boysen2021last}, which can have significant impact on their performance~\cite{bates2018transforming}. The driver decision is inherently difficult to model, as it is based on prior knowledge of specific urban areas, acquired through time and experience. Moreover, traffic and road network conditions or temporary access restrictions might make the modeling task even more challenging~\cite{nguye2019optimising}.

While optimization-based approaches have made significant strides in route planning, there remains an opportunity to leverage machine learning to bridge the gap between theoretical plans and real-life execution. By considering characteristics that describe urban context, such as building and road types, and vehicle characteristics, machine learning models have the potential to provide more accurate predictions of delivery efficiency.

The nimbleness of cargo-bikes presents a distinct advantage in dense urban environments, allowing them to find parking more quickly and closer to the final destination. However, this competitive edge is often overlooked in performance comparison studies due to the scarcity of literature and data on service times. By focusing on this crucial aspect of delivery operations, we aim to shed light on the real-world benefits of cargo-bikes.
 
Our work represents a first step towards the broader goal of accurately modeling delivery vehicle behavior across diverse urban contexts. By leveraging machine learning and rich datasets, we can develop models that better predict vehicle performance, delivery duration, and estimated delivery times.

\section{Urban delivery datasets}
Data collection challenges and private sector restrictions on data sharing in urban logistics make it a domain that is particularly difficult to study~\cite{ghizzawi2024modelling}. Datasets that capture real-world driver behaviour and delivery efficiency can provide insights that could assist operators and inform urban planning policies
Indeed, richer and larger datasets on urban logistics are a critical step to help improve urban freight and reduce the environmental footprint of transport.

\begin{table}[h!]
\centering
\caption{Comparison of Cargo Bike and Van Delivery Datasets}
\begin{tabular}{l|l|l|l}
\toprule
 & \textbf{Amazon Vans} & \textbf{Pedal Me} & \textbf{Urbike} \\
\midrule
\textit{Location} & 4 US cities & London, UK & Brussels, Belgium \\
\textit{Vehicle Type} & Delivery Vans & Cargo Bikes & Cargo Bikes \\
\textit{Delivery Type} & Primarily parcels & Mixed urban logistics & Mixed urban logistics \\
\textit{Delivery Area} & City-wide & 9-mile radius & 8 km radius\\
\textit{Parking Location Data} & Approximate geoloc & h3 hexagon & h3 hexagon \\
\textit{Distance to door} & Not available & Available & Available \\
\textit{Parcel Details} & Available & Not available & Not available \\
\textit{Service Time Calculation} & Amazon model & GPS traces & GPS traces \\
\textit{Number of Routes} & 1,131 & 1,485 & 786 \\
\textit{Number of Deliveries} & 150,657 & 11,828 & 10,965 \\
\bottomrule
\end{tabular}
\label{table:dataset_comparison}
\end{table}

In this paper, we present three datasets, including two novel cargo-bike delivery datasets. We present an overview of these datasets in Table~\ref{table:dataset_comparison}. We start by studying an Amazon van delivery dataset spanning five US cities and present some insights of delivery driver behaviour in cities. We then introduce the two new cargo-bike delivery datasets: one by Pedal Me, a cargo-bike operator in London, and one by Urbike, a cargo-bike operator in Brussels. 

\subsection{Van deliveries in US cities: Amazon Last Mile Routing Research Challenge} 
\label{sec:Amazon_Data}
\subsubsection{Amazon dataset overview}

Amazon recently open-sourced a large dataset of van-deliveries for their Last Mile Routing Research Challenge, organised in 2021~\cite{almrcc, amazon_dataset}. The challenge aimed to model sequences of delivery stops to predict the actual delivery routes conducted by experienced drivers, to ``\textit{reflect the tacit knowledge of seasoned drivers gleaned through years of experience}''.  

The Amazon dataset is one of the largest and most up-to-date van delivery datasets openly available, spanning five U.S. cities, namely Austin, Boston, Chicago, Los Angeles and Seattle. The 9,184 routes in the dataset provide good coverage of city topography, including high-density multi-family deliveries, lower density suburban deliveries, and deliveries to businesses and business centres. Each route is characterized by a variety of route-level, stop-level, and package-level features, which are described and explained in depth in the paper introducing the dataset~\cite{amazon_dataset}. The routes summarize drivers' entire work days and provide the latitude and longitude of stop locations, as well as the number of packages delivered at each stop, the dimensions of those packages, the planned service time, and the delivery status. 

\subsubsection{Descriptive statistics of the Amazon dataset}

While Merchan et al.~\cite{amazon_dataset} give an overview of the dataset, and some descriptive statistics of the routes, we delve in more details on the efficiency of deliveries. In this section, we investigate the distribution of durations for journeys between stops and the time spent servicing each stop. Table \ref{table:day_summary} provides a snapshot of an average day for an Amazon delivery driver across four cities represented in the dataset~\footnote{We have excluded Los Angeles due to OpenStreetMap tagging issues, described further in the Appendix~\ref{appendix:OSM}}. The analysis is confined to routes where all deliveries were made within the city limits. This filtering process reduces the size of the Amazon dataset from 9,184 rounds to 1,131. In the resultant filtered dataset, a driver typically spends around 2.5 hours in transit between stops and an additional 5.0 hours walking to the address and executing the package delivery. This culminates in an average total workday of 8.4 hours, or 7.5 hours when the stem mileage, i.e., the trips to and from the depot, is excluded. In contrast, when filtering for trips that occur entirely outside the city limits, the total travel time increases to 2.6 hours, while the service time decreases to 4.5 hours. This suggests that deliveries outside city limits necessitate more travel time between stops and have shorter service times.

\begin{table}[h]
\caption{Driver Round Summary Statistics for Amazon Deliveries}
\begin{tabular}{lrr|rrrr}
            \hline
                 & \multicolumn{1}{l}{} & \multicolumn{1}{c|}{\textbf{}} & \multicolumn{4}{c}{\textbf{Mean (hr)}} \\ \hline
\textbf{City} &
  \textbf{Route \#} &
  \textbf{\begin{tabular}[c]{@{}r@{}}Package \#\end{tabular}} &
  \textbf{Duration} &
  \textbf{\begin{tabular}[c]{@{}r@{}}Travel \\ Time\end{tabular}} &
  \textbf{\begin{tabular}[c]{@{}r@{}}Service \\ Time\end{tabular}} & \textbf{\begin{tabular}[c]{@{}r@{}}Stem \\ Time\end{tabular}} \\ \hline
Austin, USA &   157 & 237 & 7.5 & 2.3 & 4.6 & 0.7 \\
Boston, USA &   174 & 219 & 9.0 & 3.0 & 5.1 & 0.9 \\
Chicago, USA &  293 & 244 & 9.5 & 3.8 & 4.8 & 0.8 \\
Seattle, USA &  507 & 213 & 7.8 & 1.6 & 5.3 & 0.9 \\
\midrule
Total/Average &              1131 & 226 & 8.4 & 2.5 & 5.0 & 0.9 \\

\end{tabular}
\label{table:day_summary}
\end{table}



Compared to Table~\ref{table:day_summary}'s summary of the average drivers entire day, Table~\ref{table:travel_times} explores only the time traveling between consecutive stops inside of the city limits\footnote{Stem-mileage, i.e., travel times to and from the depot, are excluded.\label{fn:stem}}. On average, drivers spend 80 seconds driving between deliveries ($SD = 74s$), indicating a dense pattern of Amazon deliveries with relatively short distances between stops.

\begin{table}[h!]
    \caption{Summary of Amazon Travel Times between Stops}
    \begin{tabular}{lrrrrrrr}
    \toprule
    \textbf{City} & \textbf{Stop Count} & \textbf{Mean (s)} & \textbf{Std. Dev. (s)} & $\mathbf{P_{10\%}}$ & $\mathbf{P_{50\%}}$ & $\mathbf{P_{90\%}}$\\
    \midrule
    Austin, USA & 29,879 & 62 & 68 & 11 & 44 & 129 \\
    Boston, USA & 21,495 & 107 & 86 & 15 & 90 & 225 \\
    Chicago, USA & 40,306 & 114 & 82 & 16 & 111 & 219 \\
    Seattle, USA & 56,441 & 55 & 51 & 12 & 42 & 113 \\
    \midrule
     & 148,121 & 80 & 74 & 13 & 56 & 182 \\
    \end{tabular}
    \label{table:travel_times}
\end{table}

Travel time, however, only comprises 33\% of delivery drivers' days in the filtered dataset~\footref{fn:stem}. The other 67\% of their time is spent either looking for parking or walking to make deliveries, averaging 1.8 minutes of service time per stop when considering all deliveries~\cite{almrcc} and 2.5 minutes when only considering deliveries inside of city limits. 

The planned (or predicted) service time is based on a complex black-box model by Amazon. It is given on a per-package basis, which we sum over the number of packages at a stop to find the total service time. Per Amazon's data description, it includes time required to park and hand-off the package at the drop-off location~\footnote{\url{https://github.com/MIT-CAVE/rc-cli/blob/main/templates/data_structures.md}}.

The service time inside of city limits is summarized in Table~\ref{table:service_times}. Similar to the distribution of travel time, the distribution of service time also exhibits skewness, with a substantial 73.7\% of deliveries demonstrating service times below the mean within city boundaries. Variability between cities is also notable, as evidenced by the median service time in Austin, Texas, which is 45 seconds less than that of Seattle, Washington ($U = 22.2E7$, $p \leq 0.01$). The difference between cities can be explained, at least partly, by differences in urban context. 

The stops with maximum service times can be largely explained through being outliers in the number of packages per delivery. The average number of packages for deliveries with service times greater than the 99th percentile (815 seconds) is 8.3, and 88\% of the outliers have more packages than the total mean of 1.9 packages per delivery. Because of the anonymization in the the Amazon dataset, it does not differentiate between multiple packages to a single address vs. multiple packages to multiple addresses at the same stop (e.g. multiple apartments in the same building complex). From visual inspection, the outlier deliveries typically occur in the vicinity of apartment buildings, where the delivery driver might have to make many trips to and from residents' doors. 

\begin{table}[h!]
    \caption{Descriptive Service Time Statistics per City for Amazon Van Deliveries}
    \begin{tabular}{lrrrrrrr}
    \toprule
    \textbf{City} & \textbf{Count} & \textbf{Mean (s)} & \textbf{Std. Dev. (s)} & $\mathbf{P_{10\%}}$ & $\mathbf{P_{50\%}}$ & $\mathbf{P_{90\%}}$ & \textbf{Max (s)}\\
    \midrule
        Austin, USA & 30,482 & 115 & 131 & 35 & 75 & 232 & 3342 \\
        Boston, USA & 22,015 & 178 & 220 & 44 & 113 & 365 & 7325 \\
        Chicago, USA & 41,074 & 144 & 147 & 40 & 97 & 304 & 2993 \\
        Seattle, USA & 57,086 & 172 & 181 & 54 & 119 & 339 & 5637 \\
        \midrule
         & 150,657 & 154 & 172 & 42 & 102 & 314 & 7325
    \end{tabular}
    \label{table:service_times}
\end{table}

The Amazon dataset provides a valuable insight into the current state of van-delivery logistics in the U.S., revealing substantial variations in travel and service times across and within cities. These variations underline the importance of considering local conditions when planning delivery routes, specifically the features of a micro-region within a city, such as land use, road types, building densities, amenities, businesses, that affect delivery operations like finding parking, navigating to destinations, locating addresses. Moreover, the significant portion of the delivery process spent on service, rather than travel, highlights the need for more efficient delivery mechanisms and potential areas where cargo-bikes could offer advantages. We explore this potential by first looking at two cargo-bike datasets in the next section, before delving into modelling and comparisons of cargo-bike and van service time performance in urban delivery scenarios.

\subsection{Cargo Bike Deliveries in London and Brussels: Pedal Me and Urbike datasets} 
In this section, we present an overview of datasets for two cargo-bike logistics operators, Pedal Me, in London, and Urbike, in Brussels. For both data-sets, GPS trackers were installed on vehicles in their fleets. Each GPS device records the rider position and speed every 10 seconds while the rider is moving. In both cases, we matched these GPS traces with data from their Transport Management System (TMS) to extract service times. Here, the data-sets we collected and analysed were significantly more fine-grained than the Amazon data-sets regarding the position and duration of events, however, the data regarding the type of delivery or the size and number of parcels was non-existent or poor. 

\subsubsection{Pedal Me dataset overview} 
Pedal Me is a cargo-bike operator in London, with a fleet size of around 100 bikes. At the time of this study, they offer a cargo and passenger bikes service, operating within a 9-mile radius of Central London. The company services a diverse range of jobs, including multi-drop deliveries and same-day courier-style deliveries. 

The Pedal Me bikes were outfitted with 66 GPS devices and recorded from June 1st, 2021 to May 31st, 2022.  The riders' tasks are recorded in a centralised database, which includes information on delivery type, address, drop-off instructions, time-window, etc. 

Pedal Me delivers food, packages, and also does parcel pickup. To focus on operation similar to that of the Amazon dataset, we have filtered out all pickups and courier-style deliveries, as well as deliveries performed with a trailer, as Pedal Me tends to use these for larger item jobs. We also remove jobs that are ``one-off'', instead keeping only deliveries that are part of a regularly scheduled service with the customer. In total, there are 1485 unique routes for a total of 11,828 parcel deliveries. In the Pedal Me dataset, we use the GPS trace to compute the service time for each delivery. We analyze the temporal difference between the time when a bike comes to a halt prior to a delivery and the time when the vehicle starts moving again. 

\subsubsection{Urbike dataset overview} Urbike, a logistics company in Belgium, operates a fleet of 30 cargo bikes, handling a wide range of goods from parcels and food items to medications and flowers. The majority of their operations cater to larger clients and fall within the multi-drop categories. Deliveries are made within a radius of 8 km, encompassing a significant portion of Brussels. For data collection, Urbike's bikes were equipped with seven GPS devices, recording data from October 29th, 2022, to September 30th, 2023, using the Traccar GPS platform. Alongside the GPS data, Urbike maintains a booking table with delivery information. However, unlike the Pedal Me data, Urbike's delivery locations are not geofenced. To estimate service time, like in the Pedal Me dataset, we use the non-moving time at a delivery stop.  The dataset, comprising 10,965 stops made over 786 routes\footnote{Route here refers to a unique device-day pair}, provides a comprehensive view of Urbike's operations.


\subsubsection{Analysing distance to the door and cruising for parking in cargo-bike datasets}


In this section, we analyse the parking behaviour of cargo-bike delivery riders. In Section~\ref{related-works}, we presented literature that underscored substantial impact of parking on the efficiency of  urban commercial vehicle operations. However, these studies have only focused on van deliveries. For instance, \citeauthor{dalla2020commercial} revealed that commercial vehicles in Seattle spend a notable portion of their journey time, approximately 28\%, cruising for parking \cite{dalla2020commercial}. This behaviour not only affects the efficiency of deliveries but also contributes to urban congestion and pollution. 

We analyze the parking behavior of cargo-bike deliverers by using the GPS trace data. We examine where each vehicle stops and measure the Haversine distance to the address's geolocation. Contrary to the substantial cruising times for parking observed in larger commercial vehicles, such as those detailed in \cite{dalla2020commercial}, our findings for cargo-bikes Brussels and London show markedly shorter distances (see Table \ref{table:Cargo_Bike_Parking_Times}). The median parking distance is around 20 meters. This indicates that cruising for parking is negligible for cargo-bikes given that they consistently are able to park right by the delivery address (20 meters corresponds to $\approx 15$ seconds, if we assume a walking speed of 5km/h). This suggests a distinct operational advantage with minimal parking search time. These findings indicate that service times primarily involve navigating within buildings or waiting at doors, rather than cruising for parking, which potentially translates into more efficient delivery operations.

\begin{table}[h]
    \caption{Descriptive Parking Distance Statistics per City for Cargo Bikes Deliveries}
    \begin{tabular}{lrrrrrrr}
    \toprule
    \textbf{City} & \textbf{Count} & \textbf{Mean (m)} & \textbf{Std. Dev. (m)} & $\mathbf{P_{10\%}}$ & $\mathbf{P_{50\%}}$ & $\mathbf{P_{90\%}}$ & \textbf{Max (m)}\\
    \midrule
    Brussels, BE & 12072 & 31.5 & 31.2 & 10.1 & 22.1 & 61.4 & 250.2 \\
    London, UK & 9592 & 29.1 & 25.2 & 10.3 & 21.4 & 60.5 & 749.8 \\
    \bottomrule
    \end{tabular}
    \label{table:Cargo_Bike_Parking_Times}
\end{table}


\subsubsection{Comparison of Urbike and Pedal Me service-time datasets}

We have seen that cruising for parking and distance to the door are both negligible for cargo-bike operations. We now look into the service time distributions of both Pedal Me and Urbike cargo-bike riders, as summarized in Table~\ref{table:London_Service_Times}. Despite the differences in geographic location, both London and Brussels datasets exhibit a right-skewed distribution, indicating a higher frequency of shorter service times and a longer tail of infrequent, longer service times.

\begin{table}[h]
    \caption{Descriptive Service Time Statistics per City for Cargo Bikes Deliveries}
    \begin{tabular}{lrrrrrrr}
    \toprule
    \textbf{City} & \textbf{Count} & \textbf{Mean (s)} & \textbf{Std. Dev. (s)} & $\mathbf{P_{10\%}}$ & $\mathbf{P_{50\%}}$ & $\mathbf{P_{90\%}}$ & \textbf{Max (s)}\\
    \midrule
    
    Brussels, Belgium & 12072 & 246 & 126 & 114 & 223 & 417 & 1813 \\
    London, UK & 9592 & 242 & 154 & 95 & 210 & 429 & 2407 \\
    \midrule
    \end{tabular}
    \label{table:London_Service_Times}
\end{table}


In examining the service times of Urbike and Pedal Me, it becomes apparent that despite the differences in their geographical locations – Brussels and London respectively – their operational efficiencies are remarkably similar. Both Urbike and Pedal Me showcase median service times just above the three and a half minute threshold, with Urbike at 223 seconds and Pedal Me at 210 seconds ($U=61545841, p < 0.001$). 

This consistency in variability suggests that both services encounter a relatively uniform range of scenarios. Moreover, the 90th percentile values, pointing to longer delivery times of around 7 minutes for some orders, may represent the more complex delivery scenarios. It's also important to acknowledge, especially in light of our findings regarding the very consistently very short distances to the delivery address, that both Urbike and Pedal Me handle a range of deliveries, including those with specific business requirements, which may differ from the more homogenized consumer package deliveries typical of services like Amazon. 

When looking at journey times for cargo bike operations in Brussels (Urbike) and London (Pedal Me), we see they are longer and more variable than Amazon's van deliveries, pointing at less dense patterns of deliveries ($U = 9.7E7$, $p < 0.01$). Amazon's vans have significantly shorter and more consistent travel times between stops, reflecting much denser operations.

\begin{table}[h]
    \centering
    \caption{Descriptive Travel Time Statistics for Cargo Bikes Deliveries}
    \begin{tabular}{lrrrrrrr}
    \toprule
    \textbf{City} & \textbf{Count} & \textbf{Mean (s)} & \textbf{Std. Dev. (s)} & $\mathbf{P_{10\%}}$ & $\mathbf{P_{50\%}}$ & $\mathbf{P_{90\%}}$ & \textbf{Max (s)}\\
    \midrule
    Brussels, BE & 2144 & 262 & 201 & 70 & 208 & 514 & 1659 \\
    London, UK & 2580 & 323 & 278 & 74 & 250 & 666 & 2184 \\
    \midrule
    
    \end{tabular}
    \label{table:cargobike_Travel_Times}
\end{table}

\subsubsection{Discussion of cargo-bike and van data sets}\label{title:discussion-van-cargobike}

\paragraph{Overview of variation in service times across data sets}

In our analysis, Amazon vans exhibited an average service time of 154 seconds across cities. The median service time for vans was 102 seconds, lower than that of cargo bikes, with Urbike at 223 seconds and Pedal Me at 210 seconds. This indicates a generally faster service for most van deliveries. Additionally, vans showed a lower 10th percentile service time of 42 seconds, compared to 114 seconds for Urbike and 95 seconds for Pedal Me. However, the standard deviations for vans were higher, notably 220 seconds in Boston and 181 seconds in Seattle, suggesting a greater variability in van delivery scenarios. This is in contrast to cargo bikes, with standard deviations of 126 seconds in Brussels and 154 seconds in London. These statistical findings reflect different patterns of deliveries both for vans and cargo bikes, with cargo bikes exhibiting a more consistent service time within each city. 

\paragraph{Distinct Operational Profiles: Parcel Deliveries vs. Diverse Urban Logistics}

When examining the cargo-bike and van delivery datasets, it is essential to highlight the contrasting operational profiles inherent in each. The Amazon van dataset, focused on parcel deliveries, primarily to residential customers, represents a delivery pattern that is generally denser and features more consistent travel times between stops. This indicates a high concentration of deliveries within smaller areas, typical of parcel delivery services. Conversely, the Urbike and Pedal Me cargo-bike datasets, pertaining to urban logistics operations, showcase less dense delivery patterns. This is evidenced by the longer and more variable journey times for cargo bikes, reflecting a broader range of delivery types, including numerous business deliveries with distinct operational requirements. These differences in delivery density and variety between the two datasets are critical in understanding the unique challenges and efficiencies of each delivery mode.

\paragraph{Limitations in data granularity}

Another key aspect to consider is the granularity of data, especially in the context of van deliveries. The Amazon dataset, while extensive, lacks specific details on crucial operational aspects. For instance, it does not explicitly indicate where the van parked or the precise walking behavior of the drivers. This ambiguity in data can lead to uncertainties, such as the time spent searching for parking or walking distances to the delivery point. We assume from the data, that a stop in the Amazon data implies a location where the van parked, and the service time will include walking to an unknown number of addresses (while the number of parcels is known). Conversely, neither of the cargo-bike datasets provided detailed information about parcels or about the type of delivery.

\paragraph{Conclusion: Recognizing the Diversity of Urban Delivery Operations}

In light of these insights, it becomes clear that a direct comparison between the cargo-bike and van datasets may not yield fully accurate reflections of their respective operational realities. The differences in delivery density and the known types of deliveries executed by all three operators point to distinct operational contexts. However, both datasets still offer some novel insights and perspectives on the respective nature of urban delivery operations across datasets, and on the behaviour of both vans and cargo-bikes across different urban areas. 

The high variance across delivery service times also highlight strong patterns of disparate delivery situations, which lead us to explore the impact of urban context on delivery performance.
Many of the deliveries in the Amazon dataset were done outside of the dense city centre, unlike the cargo-bike operations, that are more focused in their respective city centres. These differences may lead to a skew towards faster suburban deliveries in the Amazon dataset in the distribution of service times when compared to the urban deliveries. Indeed, we hypothesize that dense urban areas, potentially due to factors like parking and navigation difficulties, may lead to significant challenges for deliveries, which would result in longer service times than in e.g. suburban or residential areas. We explore this hypothesis through the design of models to predict the service times of different vehicle types across diverse urban areas.

\section{Urban context}
\label{sec:models-of-urban-context}
\vspace{-2mm}

In the previous section, we highlighted the important variation in vehicle performance in cities. Specifically, we hypothesised that the type of urban area where a delivery takes place has a strong influence in the efficiency of a delivery. In this section, we introduce the challenge of modelling the performance of delivery vehicles (here, vans and cargo-bikes) across different urban contexts. 
\vspace{-1mm}
\subsection{Dividing cities in micro-regions}
\vspace{-1mm}
Cities, with their vast diversity, demonstrate both unique and shared geometric characteristics across their spatial forms. The field of urban morphology, rooted in geography and architecture, has looked into the patterns and types that exist within urban constructs, from individual buildings to overarching street patterns~\cite{fleischmann2022methodological}. Works by researchers like Steadman~\cite{steadman2000classification} and Marshall~\cite{marshall2005urban}, or more recently Boeing~\cite{boeing2019urban}, have critically analyzed the geometries of urban elements, leading to a better understanding of urban configurations. As the world grapples with pressing challenges, there's a heightened need for evidence-based urban studies~\cite{gramacki2021gtfs2vec,sefala2021constructing,wagner2022using,dong2019predicting,joshi2022identifying}. 

In essence, urban morphology provides a lens to view and analyze the urban tissue of cities. In our case, the challenges faced by a delivery driver—such as finding appropriate parking, navigating to the delivery spot, and pinpointing an exact apartment or room within a complex—all relate to how the surrounding urban elements influence how a delivery happens. We call this the `urban context` of a delivery.

In this section, we introduce our framework to capture these urban contexts in a principled way, aiming to capture the statistical patterns across urban areas, to then use it to study vehicle delivery performance.

Exploring urban contexts predicates a method to divide the urban space into regions. Common options include context-aware micro-regions such as census zones, zip codes or neighborhood boundaries. These methods have the downside of requiring some manual work to retrieve them, and the rules behind these boundaries can vary widely across cities and countries, making them difficult to compare. To achieve a consistent and efficient representation of urban spaces, we adopt a modern approach. Departing from traditional, often labor-intensive methods, we leverage Uber's Hexagonal Hierarchical Spatial Index (H3) as recommended by Wozniak et al. \cite{wozniak_hex2vec_2021}. This enables us to divide the delivery area into micro-regions, providing a uniform and adaptable tessellation irrespective of city or country. Uber H3 is a global grid system that partitions the world into hexagons of various sizes, making geospatial analysis more efficient and precise. The Uber H3 system is hierarchical and has 16 levels of cell resolutions. We choose the H3 resolution of 9 (average edge length of 174.4 meters), as it captures several city blocks effectively~\cite{wozniak_hex2vec_2021}. 

\begin{figure}[h!]
    \centering
    \includegraphics[width=\textwidth]{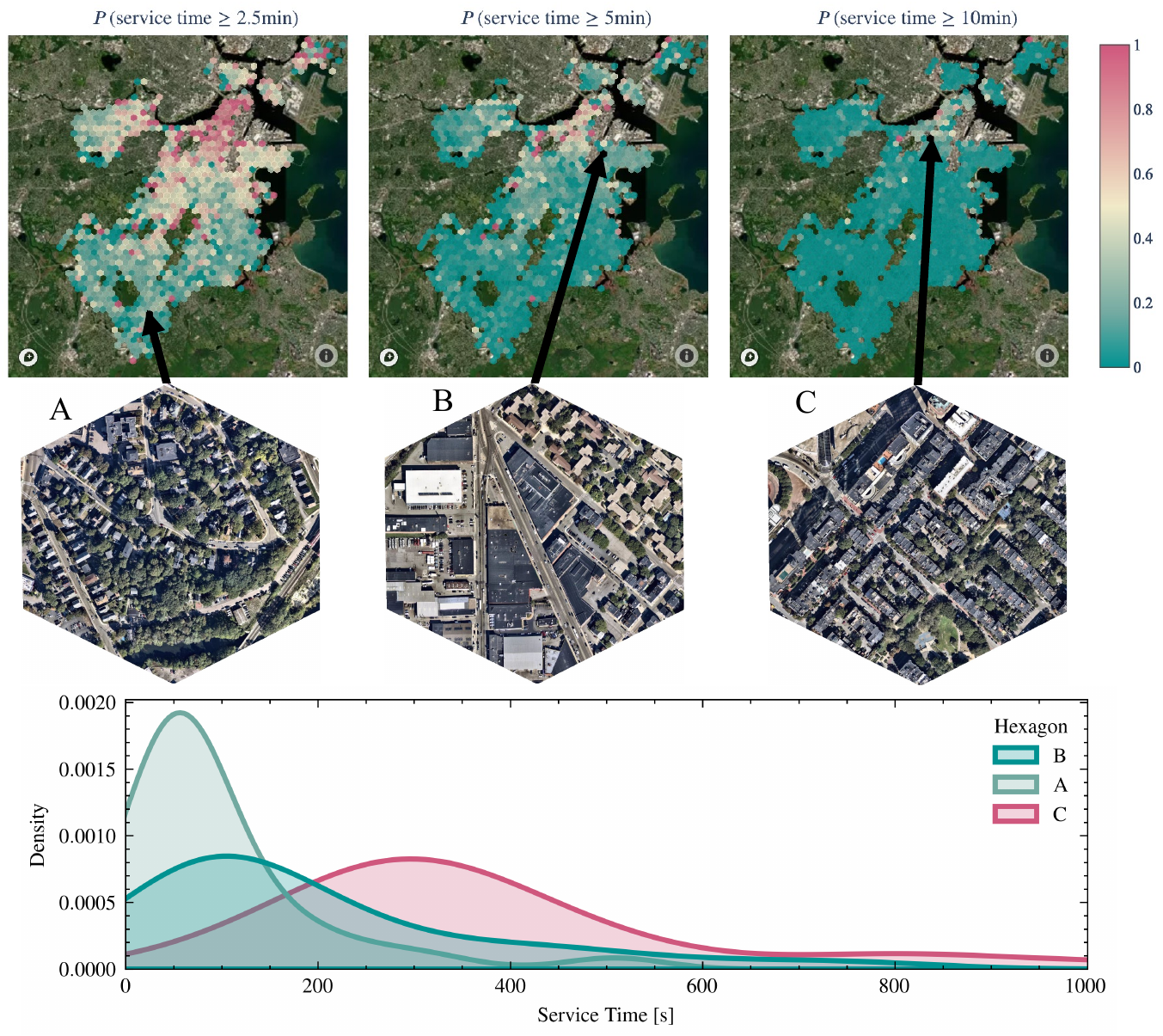}
    \caption{The probability of delivery service time within a hexagon lasting longer than 2.5, 5, and 10 minutes respectively in the Amazon van dataset. We show three hexagons (A-C) sampled from three different areas in Boston to illustrate the effect of urban context on service time. A has a median service time near the 5th quantile and is located in a residential neighborhood. B has a service time close to median and is located in a mixed use area. C, a hexagon with an expected service time near the 95th quantile, is in the Boston downtown.}
    \label{fig:urban_context}
\end{figure}

Figure~\ref{fig:urban_context} employs H3 gridding to represent service durations in Boston, using data from the Amazon dataset. It illustrates three maps, each depicting the likelihood of delivery service times within individual hexagons exceeding 2.5, 5, and 10 minutes, respectively. 

Downtown Boston, depicted in the upper right region of the plot, serves as a striking focal point, given its dense urban context and notably longer service times. This is a testament to the role urban density plays in the delivery dynamics, clearly illuminating the correlation between service time and urban context.

The hexagons showing satellite imagery of three distinct urban micro-regions reveal a range from residential to urban delivery areas, underlining the heterogeneous nature of the micro-regions. For example, Hexagon A, exclusively containing single-family homes in a residential area, displays the shortest delivery times. The swift service can be attributed to readily available parking adjacent to each home, thereby minimizing walking distances.

The distribution of service times illustrates a heavy tail, with a few specific hexagons in the city center accounting for the majority of deliveries exceeding 10 minutes, while most hexagons on the outskirts tend to have a majority of delivery service times under 2.5 minutes. The variability in service times across the different regions underscores the importance of taking into account when modelling delivery performance.

Hexagon C encompasses two zip codes within Boston's densely-populated Back Bay neighborhood, which had a population density of $9706\:\text{people} / \text{km}^2$ as per the 2010 US Census~\cite{boston_census}. Hexagon C has the highest median service time in Boston, meaning Amazon delivery drivers can expect to have the longest service times in the city when they deliver to Hexagon C. Interestingly, Hexagon C falls in the 53rd percentile of buildings per hexagon for the city of Boston, indicating that service times, while influenced by urban context, are affected by a diverse array of factors beyond just building count.

We propose a numerical representation of urban context that can serve as a component in modelling service time, which we describe in the following sections.

\subsection{Embedding urban micro-regions}\label{sec:OSM-probs}

\subsubsection{Representing entities as vector embeddings}

In computational science, a key challenge lies in how to represent, or "embed", complex entities in a way that captures their similarities yet is also manageable for computational tasks. Conceptually, an embedding is a mapping – it's possible to map data from a discrete space, such as words in a language or tags in a map, to a continuous vector space based on their similarity. Historically, embeddings gained prominence in the field of Natural Language Processing~\cite{mikolov2013linguistic, pennington2014glove, devlin2018bert}. 


When trying to represent the characteristics that define an urban area, the challenges are reminiscent of those in language. Urban micro-regions, like words, are complex entities with numerous attributes – road types, building densities, amenities, and so on. To work with such varied data efficiently, we can "embed" these regions in a manner similar to words. By transforming our urban data into a continuous vector space, we create representations where similar urban contexts are close together in the embedded space. Embeddings facilitate dimensionality reduction, compressing this information into lower-dimensional vectors while retaining essential characteristics. This is particularly crucial given the complexity and variety of urban features. Additionally, embeddings can capture the non-linear relationships and context-dependent similarities between urban micro-regions, a subtlety that raw feature counts might oversimplify. 

\subsubsection{Using OSM tags as urban descriptors}
To capture the characteristics of urban micro-regions, we use OpenStreetMap (OSM), an open-source mapping project~\footnote{\url{ https://www.openstreetmap.org }}. OSM is considered a leading example of volunteered geographic information~\cite{mashhadi2015impact}, having over 9 million users and approximately 4 million daily map edits as of December 2022~\footnote{\url{https://wiki.openstreetmap.org/wiki/Stats}}. OSM has become a popular source of geospatial data in recent years, with its uses spanning many disciplines. The data is widely available for download through various APIs and provides rich spatial and semantic information, such as building footprints and types.  

In our work, the urban context for each micro-region is based on the types of buildings, plots, and roads annotated in OSM in that region.
OSM tags include information about businesses, roads, natural features, amenities, and more. 
While OSM is a valuable open-source geographic dataset, its quality and completeness can vary significantly across regions~\cite{haklay2010good, hecht2013measuring}. Urban areas generally have better coverage and data quality compared to rural areas, however, inconsistencies in tagging can still pose challenges~\cite{vandecasteele2015improving, helbich2012comparative} (we present more details about the OSM dataset in Appendix~\ref{appendix:OSM}). Recent methods based on machine learning have been proposed to improve the quality and coverage of OSM data \cite{vargas2020openstreetmap}. Moreover, the representation learning techniques we are applying abstract away from tags, creating embeddings that are less affected by annotation accuracy than simple tag counts.


\subsubsection{Leveraging embedding models for urban micro-regions}
\label{sec:geovex-and-geotags}
Embeddings using OSM features, such as GeoVectors~\cite{tempelmeier2021geovectors} and Hex2Vec~\cite{wozniak_hex2vec_2021}, have been shown to exhibit strong semantically interpretable characteristics of urban space. 


In this study, we used the \emph{srai} library\footnote{See the \emph{srai} library at \url{https://github.com/kraina-ai/srai}}, a versatile Python framework for geospatial data handling and analysis, as our foundational platform. \emph{srai} is distinguished by its ability to download geospatial data, segment areas into micro-regions using various algorithms, and train embedding models with diverse architectures, as detailed in the work by Gramacki et al.~\cite{gramacki2023srai}. 

Within this framework, we implemented the GeoVex model~\cite{donghi2023geovex}\footnote{An example of GeoVex being used in \emph{srai}: \url{https://kraina-ai.github.io/srai/0.6.2/examples/embedders/geovex_embedder/}}, a state-of-the-art geospatial location model, adapting it to enhance our analysis and predictions in geospatial AI applications. While its precursor Hex2Vec \cite{wozniak_hex2vec_2021} learns an embedding for each hexagon based only on the tags present in that hexagon, GeoVex expands the input features by also including information from neighbouring hexagons (expanding the use Tobler's first law of geography: `Everything is related to everything else, but near things are more related than distant things''). This allows the model to learn embeddings that are contextualized based on the surrounding environment. GeoVex also adapts the latest advancements in convolutional autoencoders to better handle the hexagonal grid structure of the data. Specifically, it uses hexagonal convolutions that align with the grid geometry. This allows the model to learn higher quality representations. Additionally, GeoVex uses a location-weighted loss function that weights the importance of hexagons based on their distance from the target hexagon. This loss formulation aligns with the structure of geographic data distributed on a hex grid centered on a location to embed. Finally, it uses a zero-inflated Poisson reconstruction layer to adapt the learned embeddings to the sparse, zero-inflated count data in a principled way without any task-specific retraining.

While our research employs the proposed architecture of GeoVeX, we do so with modifications to the input tags and sub-tags to better reflect essential features for predicting delivery times. As a result, our model incorporated 681 OSM sub-tags. Notably, we included \texttt{highway} tags due to the relevance of the quantity and variety of roads within a hexagon for service time prediction. However, we chose to exclude tags that were deemed non-influential or problematic, such as \texttt{natural} and its sub-tags, owing to their high variability across different cities~\footnote{The full list of tags can be found at \url{https://github.com/green-last-mile/cargo-bike-analysis/blob/main/data/geovex/target_tags.txt}}. Furthermore, we restricted the training of the embedding model to cities for which we had delivery data, as expanding the training to include additional cities did not enhance the performance of the embeddings for our target cities. 

\section{Empirical performance of delivery vehicles across urban regions}
\label{sec:Clustering}

In this section, we use our model of urban context to study the empirical service times of vans and cargo-bikes across our data-sets. To be able to draw comparisons across vehicles and cities, we choose to go up in the hierarchy of scale, following the assumption that cities share a common high-level structure, namely that they can broadly be divided into an urban core, dense residential areas, commercial or industrial regions, and suburban zones. First, we begin by defining urban regions based on our model of micro-regions. We then study the resulting areas across the different cities that our data sets span.

\subsection{Categorising urban regions by clustering micro-region embeddings}


To analyze vehicle performance both within and between cities, we utilize our micro-region models to identify types of urban areas that exhibit similar characteristics. This approach helps us understand how vehicles perform in areas with comparable urban features. We follow ~\citeauthor{wozniak_hex2vec_2021}~\cite{wozniak_hex2vec_2021} and use agglomerative clustering to group hexagons with similar embedding vectors. Agglomerative clustering is a hierarchical clustering technique that incrementally constructs a tree-based structure by repeatedly merging the nearest pair of clusters, starting with individual data points as singleton clusters. This bottom-up approach is guided by a distance metric, such as Euclidean distance, which quantifies the similarity between data points or clusters. We apply this agglomerative clustering algorithm to hexagons from a single city independently. We decide to divide cities in four different clusters, as upon inspection it yielded coherent and interpretable area types across cities. 




\subsection{City descriptions through models of urban micro-regions}\label{section:urban-clusters-all-cities}


We study the urban landscapes of all six cities (Boston, Seattle, Austin, Chicago, Brussels, and London) with the aim of uncovering and comparing the underlying urban structures and functional zones of these cities in an intuitive way. We transformed raw OpenStreetMap (OSM) tags into more encompassing "super-tags" such as Built Environment, Transportation, and Leisure \& Recreation (see Appendix~\ref{Appendix:city-descriptions} for more detailed descriptions). This enabled the identification of distinct types of urban areas within each city, providing a nuanced understanding of both unique and shared characteristics. The integration of population data from Meta’s high-resolution population maps enriched our analysis, offering a detailed perspective on urban density and distribution.


 Despite geographical and cultural differences, we found common urban elements such as central urban cores, residential areas, and commercial or industrial hubs in all cities. However, each city also exhibited unique traits. Notably, Brussels and London stood out with their rich historical and cultural clusters, while the American cities demonstrated various degrees of suburban sprawl.

A key finding was the universal significance of urban cores, which, regardless of their size, emerged as pivotal areas in all cities. For example, Boston's Downtown Core, though only comprising 98 hexes, and London's Urban Center, spanning 1202 hexes, both showed high concentrations of amenities and transportation facilities. This highlights the central role of these cores in urban connectivity and activity.

In terms of residential areas, there was a marked variation. Seattle's Suburban Residential area, covering 981 hexes, sharply contrasted with Austin's Rural/Suburban cluster, the largest in our study with 1889 hexes, reflecting the diversity in urban sprawl and residential planning. Additionally, unique clusters were identified in specific cities. Brussels, for instance, featured a "Cultural Hub" with the highest population density (1254.58 per cluster) and significant historical and cultural elements, a pattern not observed in the American cities.

The distribution of super-tags provided further insights into urban dynamics. Chicago's Industrial Hub, for instance, had the highest Built Environment index (165.81), indicating its industrial emphasis. In contrast, London's Cultural Hub exhibited a high Commerce \& Industry index (11.86), mirroring its fusion of cultural richness and commercial activity.

In conclusion, this comparative analysis of urban clusters across Boston, Seattle, Austin, Chicago, Brussels, and London, sheds light on both the shared and unique aspects of urban planning. While there is a foundational similarity in the high-level structure of these cities - each possessing an urban core, dense residential areas, commercial or industrial regions, and suburban zones - the nuances within each cluster reveal the distinct identity and function of the urban areas.
For instance, the "Downtown Core" in one city may vary significantly in its cultural and historical richness compared to another. Similarly, suburban areas across different cities, while sharing residential characteristics, differ in their population density and amenities. These variations underscore the importance of nuanced analysis when comparing urban regions across cities. Despite these differences, the presence of shared functional zones establishes a basis for comparison and highlights the underlying common urban structures. This reinforces the idea that, while cities are unique, their urban compositions may mirror each other, allowing for meaningful cross-city comparisons and insights.

\subsection{Empirical service times of vans and cargo bikes across urban regions}
\label{sec:service_times_comparison}

\begin{figure}
    \centering
    \includegraphics[width=\textwidth]{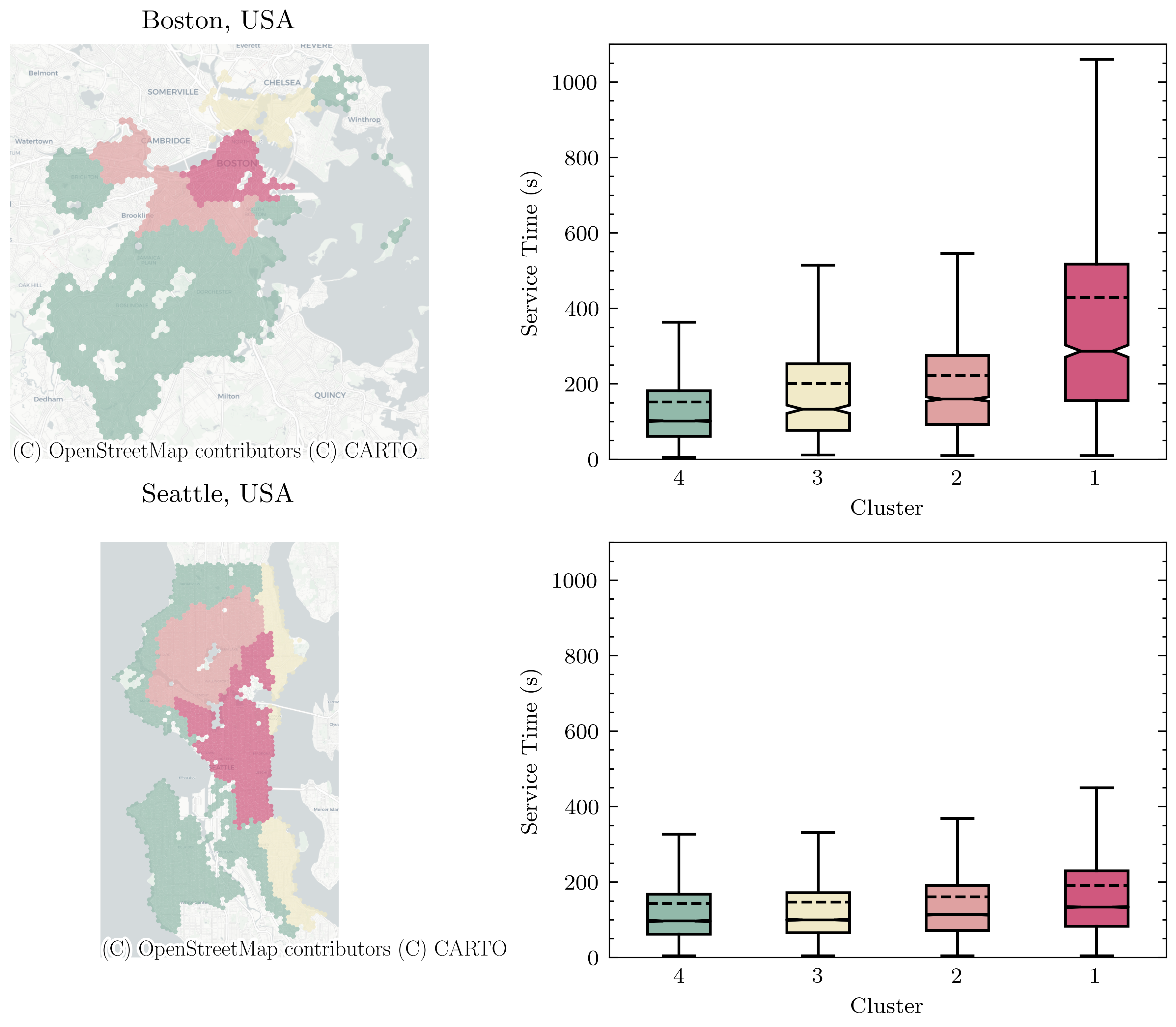}
    \caption{Box plots of Amazon van service times by within-city cluster. The clusters are organized such that 1 has the highest mean service time and 4 the lowest. The top row is a summary of deliveries in Boston, Massachusetts and the bottom of Seattle, Washington.}
    \label{fig:cluster_van_times}
\end{figure}

In Section~\ref{title:discussion-van-cargobike}, we hypothesised that urban context may have a strong influence on delivery service times, and that this may affect vehicles in different ways, depending on their type. Figure~\ref{fig:cluster_van_times} confirmed this intuition, showing that in Boston, a few hexagons in the city center account for the majority of delivery service times above 10 minutes. We explore this effect further by looking at the distribution of service times across the clusters described in Section~\ref{section:urban-clusters-all-cities}. We begin by looking at the van performances across urban areas in the cities of Boston and Seattle (Figure \ref{fig:cluster_van_times}). 

The resulting distributions of service times inside of the clusters show the strong relationship between urban context and service time, as well as highlighting the heavy-tailed nature of service times. Using the Kruskal-Wallis test, we show that there is a statistically significant difference between the service time distributions for both Boston ($H(3) = 695.21, p \ll 0.001$) and Seattle ($H(3) = 497.70, p \ll 0.001$).

The Downtown Core has the highest average and median service time in both cities. As illustration, the Downtown Core (cluster 1) in Boston contains Hexagon C from Figure~\ref{fig:urban_context}. We conducted post-hoc analysis using Durbin-Conover tests with Holmes correction to compare the individual clusters. In Boston, the Downtown Core and the Outer Urban Residential (cluster 2) areas  have a significant difference ($p<0.001$) and over two minutes of difference between their median service times. The Outer Urban Residential area is also has significantly slower service times than the Mixed Use area (cluster 3), though with only 27 seconds in difference between the median service times ($p=0.00$). Most deliveries happened in the Suburban Residential area (N=17641), also by far the largest cluster in Boston, but also least densely populated (795 hexagons). This is were service times were the shortest, significantly so when compared to all other areas.

Although the median differences in service times were less marked across the clusters in Seattle, we still find significant differences across all clusters. The Downtown Core of Seattle saw a much less affected median service time than in Boston (134 seconds and 287 seconds respectively). Since the Downtown Core area captured by the clustering is much larger in Seattle than in Boston (497 vs 98 hexagons), it is plausible that a denser and smaller area would have shown a stronger effect.

The higher planned service time in the Downtown Core of both cities can be explained, in part, by longer time spent looking for parking, as well as considerable walking time between addresses instead of driving. Although not made explicit in the dataset, we believe each stop represents the vehicle being parked, and the service time will include the time looking for parking, then walking between the different addresses to deliver parcels. The Downtown Core in Seattle spans the region that was studied in a recent work, where the authors performed an empirical analysis on parcel delivery drivers and found they spend 138 seconds cruising for parking on average~\cite{dalla2020commercial}. Their findings hint that part of the increased service time is likely due to drivers spending longer looking for parking. 

\begin{figure}
    \centering
    \includegraphics[width=\textwidth]{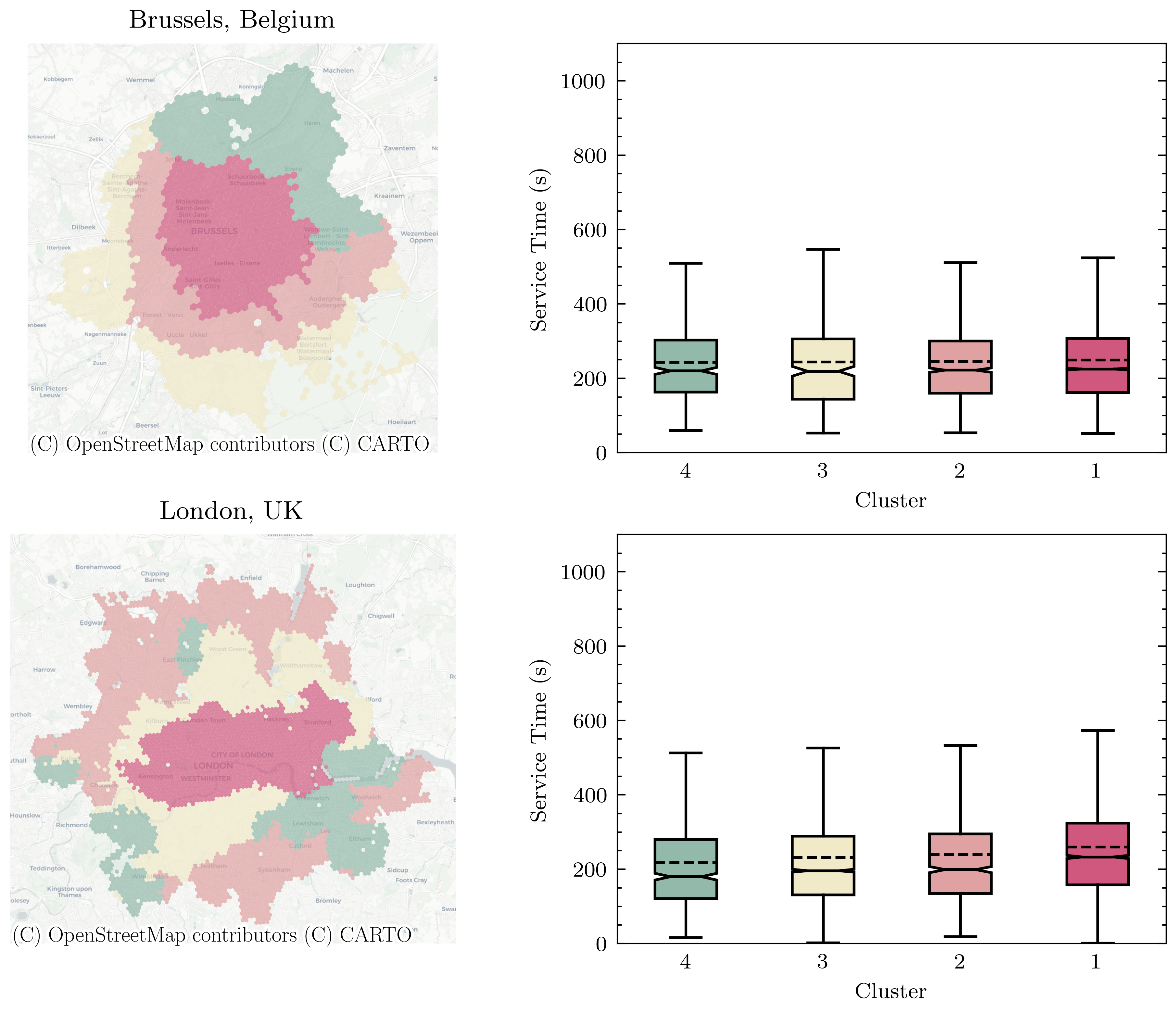}
    \caption{Box plots representing the empirical cargo-bike service times across city clusters. The clusters are organized such that 1 has the highest median service time and 4 the lowest. The top row is a summary of delivery service times in London, United Kingdom and the bottom of Brussels, Belgium.}
    \label{fig:cluster_bike_times}
\end{figure}

Next we look at the effect of urban context on the service time of cargo-bike deliveries, both in London and Brussels. We applied the same clustering methodology described in Section \ref{sec:Clustering} to London, as illustrated in Figure \ref{fig:cluster_bike_times}. Our analysis revealed a statistically significant difference in the within-cluster service times, as determined by the Kruskal-Wallis test ($H(3) = 640.10, p \ll 0.001$). The Urban Centre had significantly longer service times than the three other clusters using the post-hoc Durbin-Conover test, with a median time 37 seconds longer than the so called "Cultural Core" (cluster 3 in Figure~\ref{fig:cluster_bike_times}). There was no significant difference between the Cultural Core and the Urban Residential area ($p=0.33$), with similar median times for both (196 seconds and 199 seconds respectively).

In Brussels, we found no significant difference across urban areas under the Kruskal-Wallis test ($p=0.33$). The median times were remarkably similar, falling all between 220 and 225 seconds across the four different clusters. This is particularly striking given the variations observed across clusters both in population density and in urban context statistics (see Appendix~\ref{Appendix:city-descriptions}). One potential explanation could be the limited number of customers studied in the dataset, reflecting similar kinds of customer needs at the delivery points, as well as cargo-bikes being unaffected by the urban context for parking.

\subsection{Discussion of results}


We have so far developed a model of urban micro-regions, and also analyzed the empirical data of two distinct datasets: Amazon van deliveries in five US cities, and cargo-bike deliveries in London and Brussels. These efforts have highlighted the significant influence of urban context on van service times, in contrast to the lesser impact on cargo-bike deliveries. Building on this foundation, our next step is to design a general model that can effectively predict service times for both types of delivery vehicles across different urban settings. This model is aimed at capturing the complex relationship between urban environment and delivery efficiency, utilizing the limited but valuable data available in urban logistics.

As we highlighted in Section~\ref{title:discussion-van-cargobike}, the datasets studied represent different operational models, with the Amazon vans dataset focusing on parcel deliveries primarily in residential areas, and the cargo-bike dataset encompassing a broader spectrum of urban logistics operations.
In advancing to design a general model, we emphasize that our objective is not to compare vans and cargo-bikes directly. Rather, we aim to demonstrate that our model can accurately predict service times for each type of delivery vehicle within their respective operational contexts across the studied cities, and potentially be applied in new urban environments. This distinction is crucial, as it highlights our focus on modeling the unique characteristics and challenges of each delivery mode in varying delivery contexts.
While we believe that a direct comparison between vans and cargo bikes could be possible with more fine-grained data (i.e. by gathering data that distinguishes cruising for parking, walking, time spent at the door/with the customer, and package-level information), such an analysis is outside the scope of this current study. 
 

\section{Vehicle Specific Models of Delivery Service Times Across Urban Environments}

In this section, we extend our model of urban micro-regions to create vehicle-specific models that accurately predict service times for both vans and cargo-bikes, leveraging our previously studied datasets. Our objective is to develop models that can generalize effectively to new and unseen urban contexts, where data might be sparse or unavailable. 
In Section~\ref{sec:models-of-urban-context}, we introduced Uber's Hexagonal Hierarchical Spatial Index (H3) to divide urban areas into micro-regions, providing a consistent and adaptable framework for our analysis irrespective of city or country. We aggregated OSM tags at the H3 cell level, to take into account diverse urban features, from land-use and road types to building densities and amenities, essential for predicting service times. We then uses the GeoVex model, to not only learn from the tags within each hexagon, but also incorporate information from neighboring hexagons, using advanced techniques like hexagonal convolutions and a location-weighted loss function. Our tailored use of the GeoVex model included a careful selection of OSM sub-tags and a focused training approach on cities with available delivery data, ensuring a nuanced understanding of urban contexts for accurate service time prediction.
Here, the overarching aim of our modelling effort is to accurately capture the distribution of service times at the hexagon (or micro-region) level. 

\subsection{Baseline Models}
\label{sec:baseline_models}

To benchmark the performance of our ``context-aware'' models, we first introduce a set of baseline models. These baseline models are deliberately simplistic, designed to provide a comparative foundation by disregarding detailed urban context information such as OSM tags or geospatial embeddings. They rely on city-specific historical data trends but do not account for the intricacies of urban geography.

The first baseline model, the City-Wide Average Model, predicts service times based on the average of all deliveries within a city. Its simplicity, however, is also its limitation as it overlooks local variations in service times. Next, we introduce the K-Ring Model. The model uses the distribution of service times within a k-ring radius of the target hexagon, offering a broader perspective than the city-wide model but potentially smoothing over important local differences. These models reflect the kinds of models that operators may use in their operations to estimate service time, based on their historical customer data, either at an aggregated city-wide level, or tailored to specific customers. Both the City-Wide Average Model and the K-Ring model assume that the underlying distribution is log-normal, fitting the distributions using quantile matching~\cite{cook2010determining}.

While these naive models are useful as baselines,
they become unusable in new geographies, where no data exists.

\subsection{Urban Context Aware Predictive Models}
\label{sec:pred_models}

With the objective of modelling service times as a function of the urban context of a delivery, we explored two prediction methods: conformal prediction and probabilistic predictions with natural gradient boosting.  The urban context was represented in two different ways. The first was the use of embeddings from the GeoVex model, as detailed in Section~\ref{sec:geovex-and-geotags}. The second was the direct application of OSM sub tag counts, to evaluate the effectiveness of geospatial embeddings. The 50-dimensional GeoVex embeddings represent a significant compression of the OSM sub-tag information, where we consider 754 sub tags.

The first predictive model considered is a conformal predictive system (CPS) for regression, a recent advancement in the field of conformal predictors. Unlike regression, which yields a single prediction for a given set of features, CPS models the cumulative distribution under which the prediction falls~\cite{bostrom2021mondrian}. Given the cumulative distribution, it is also feasible to extract quantile predictions, as well as the mean and variance. The distribution-free nature of conformal methods frees the modeller from the need to choose the underlying distribution.

We specifically implement a variant of CPS, the Mondrian conformal predictive system. The Mondrian modification, partly due to its effective handling of heteroskedastic residuals, have been shown to outperform both standard and normalized conformal predictors in terms of predictive accuracy and informative output~\cite{bostrom2021mondrian}. We utilized the Crepes~\cite{crepes} Python library to implement the Mondrian CPS models, with XGBoost serving as the base regressor~\cite{chen2016xgboost}. Known for its efficiency and effectiveness in handling complex datasets, XGBoost is particularly suited for processing the relationships inherent in the urban context feature set.

Our second model utilized probabilistic modeling via natural gradient boosting, specifically leveraging the XGBoostLSS package~\cite{Maerz2023}. This package is an extension of NGBoost~\cite{duan2020ngboost}, which provides a generalized framework for probabilistic regression by considering the parameters of the conditional distribution as targets within a multi-parameter boosting algorithm. The modeller is given the flexibility to select the appropriate conditional distribution. Consequently, as the conditional distribution is predicted for each instance $x$, parameters such as the mean, median, and other desired quantities can be readily derived.

The framework we employed supports the use of any \textit{proper} scoring rule as the loss function. In probabilistic regression, a scoring rule is \textit{proper} if the true distribution of the data maximizes the expected score~\cite{gneiting2007strictly}. For our study, we have selected the negative log likelihood (NLL) as the loss function. NLL quantifies how well the probability distribution predicted by the model explains the observed data.

\subsection{Results}

Our aim in modeling service time as a function of urban context is to accurately predict the distribution of service times for a given region. Considering the operational differences between cargo bikes and vans as outlined in Section~\ref{title:discussion-van-cargobike}, we have developed distinct models for each mode of transport and present their analyses separately.

To compare the modeling approaches discussed in Section~\ref{sec:pred_models} against the baseline models from Section~\ref{sec:baseline_models}, we utilized a set of metrics. These are the mean interval coverage targeting $90\%$ with a significance level of $\alpha = 0.05$, the mean interval width, and the pinball loss at the 50th and 95th percentiles.

The pinball loss, also referred to as quantile loss, is used to measure the accuracy of quantile predictions. Defined for a quantile $\tau$, the pinball loss is given by:

\begin{equation}
L_{\tau}(y, \hat{y}) = 
\begin{cases} 
\tau (y - \hat{y}) & \text{if } y \geq \hat{y}, \\
(1 - \tau) (\hat{y} - y) & \text{if } y < \hat{y},
\end{cases}
\end{equation}

where $y$ is the true value and $\hat{y}$ is the quantile prediction. This metric penalizes overestimation and underestimation asymmetrically, thereby reflecting the cost of deviation from the desired quantile. A lower pinball loss suggests superior quantile prediction accuracy. In our evaluation, we focus on the median (50th percentile) and the upper tail (95th percentile) to analyze the central tendency and the spread of the predicted distribution, respectively. When pinball loss is evaluated at the median ($\tau = 0.5$), it is equivalent to half of the mean absolute error.

Additionally, we evaluated models using the Continuous Rank Probability Score (CRPS), which measures the accuracy of a predicted cumulative distribution function $F_{\theta}$, parameterized by $\theta$, against an observed value $y$. CRPS is calculated by the following integral:
\begin{equation}
C(\theta, y) = \int_{-\infty}^{y} F_{\theta}(z)^2 dz + \int_{y}^{\infty} (1 - F_{\theta}(z))^2 dz.
\end{equation}
CRPS assesses the divergence between the forecasted distribution and the actual observation, with a lower CRPS indicating a more accurate predictive distribution. This metric is especially valuable in our context, as it is expressed in the same units as the predicted variable, allowing for a direct interpretation of the model's performance.

Incorporating the baseline models, we assess a total of five distinct models:

\begin{itemize}
    \item 3-ring: This model is an implementation of the K-ring baseline model with three concentric rings.
    \item City: This baseline model captures city-level service time variations.
    \item CPS-Geo: The Conformal Predictive System model utilizing a Mondrian approach with XGBoost as the base regressor. The feature set is GeoVex embeddings.
    \item XGB-Geo: An XGBoostLSS model that assumes a Lognormal conditional distribution and employs GeoVex model embeddings as features for a hexagon.
    \item XGB-OSM: Another XGBoostLSS model with a Lognormal conditional distribution which, in contrast to XGB-GeoVex, uses the complete set of OSM subtags as features.
\end{itemize}

The models were implemented so as to provide both quantile and sampling based prediction. The XGBoost-based models underwent hyperparameter optimization using Optuna~\cite{akiba2019optuna} on the Boston dataset. 

\subsubsection{City-Specific Model}
\label{sec:city_model}

In Table~\ref{tab:city_van}, we present a comparative analysis of the five models across two cities, treating each city's data independently and applying 5-fold cross-validation to assess model performance. The dataset was split on the hexagon level, resulting in training and evaluation sets that had unique hexagons, avoiding contamination of the test set. The metrics used for comparison include the CRPS, mean interval coverage and width, as well as the pinball loss at the 50th percentile ($L_{0.50}$) and 95th percentile ($P_{95}$). The best value of the three predictive metrics for each metric is shown in bold.

\begin{table}[h]
    \caption{\textbf{City-Specific Van Model Comparison}. The baseline models are separated from the predictive models by a horizontal line. The best value of the baseline models in each metric is underlined, with the best value for the predictive models being bolded.}
    \label{tab:city_van}
    \begin{tabular}{lllllll}
    \toprule
     &  & \multirow[*]{2}{*}{CRPS (s)} & \multicolumn{2}{c}{Mean Interval} & \multicolumn{2}{c}{Pinball} \\
     City & Model &  & Coverage (\%) & Width (s) & $P_{50}$ (s) & $P_{95}$ (s) \\
    \midrule
    \multirow[*]{5}{*}{Boston} & 3-Ring & \underline{$78.1\pm8.5$} & \underline{$90.1\pm1.2$} & \underline{$422.6\pm28.2$} & \underline{$51.2\pm5.4$} & \underline{$28.7\pm3.3$} \\
     & City & $85.8\pm9.1$ & $91.2\pm1.0$ & $478.0\pm0.0$ & $55.2\pm5.4$ & $34.5\pm3.3$ \\
     \cline{2-7}
     & CPS-Geo & $\mathbf{78.2\pm8.1}$ & $\mathbf{90.2\pm1.0}$ & $437.8\pm26.4$ & $\mathbf{51.6\pm5.3}$ & $\mathbf{29.6\pm3.3}$ \\
     & XGB-Geo & $78.5\pm8.4$ & $82.0\pm1.8$ & $\mathbf{320.1\pm17.7}$ & $51.7\pm5.3$ & $31.3\pm4.0$ \\
     & XGB-OSM & $79.5\pm8.5$ & $87.1\pm0.9$ & $366.6\pm17.3$ & $52.4\pm5.4$ & $31.6\pm3.7$ \\
    \cline{1-7}
    \multirow[*]{5}{*}{Seattle} & 3-Ring & $71.4\pm3.3$ & \underline{$89.8\pm0.9$} & \underline{$380.1\pm11.5$} & \underline{$46.5\pm2.0$} & \underline{$26.4\pm1.5$} \\
     & City & $74.4\pm3.2$ & $88.5\pm0.6$ & $412.2\pm0.0$ & $48.3\pm1.9$ & $28.1\pm1.3$ \\
    \cline{2-7}
     & CPS-Geo & $\mathbf{70.2\pm3.1}$ & $\mathbf{88.8\pm0.7}$ & $398.9\pm10.0$ & $\mathbf{46.6\pm2.1}$ & $\mathbf{26.3\pm1.5}$ \\
     & XGB-Geo & $70.6\pm3.4$ & $84.4\pm1.4$ & $\mathbf{316.6\pm11.4}$ & $46.8\pm2.1$ & $27.4\pm1.8$ \\
     & XGB-OSM & $70.4\pm2.9$ & $87.8\pm1.3$ & $354.2\pm8.6$ & $46.8\pm1.9$ & $26.7\pm1.3$ \\
    
\bottomrule
    \end{tabular}
\end{table}

In Table~\ref{tab:city_van}, the 3-Ring baseline model stands out for its performance in Boston, achieving the best results across all metrics with a CRPS of 78.1 seconds and a pinball loss at the 50th percentile of 51.2 seconds. This model's ability to capture the localized urban context—beyond just the city level—plays a pivotal role in its predictive success, underscoring the importance of granular spatial data in understanding delivery dynamics. While the 3-Ring model performs well, it is limited by its lack of generalizability. The model's methodology, which does not rely on a conventional test-train split but instead uses all available data to create the contextual distribution, limits its applicability across different datasets and potentially overstates its performance due to not being subject to the same validation constraints as the other models. 

The CPS-Geo model was the best performing predictive model for Boston, with a CRPS nearly identical to the k-ring model at 78.2 seconds, but with a slightly higher pinball loss at the 50th percentile of 51.6 seconds. Its wider coverage interval of 437.8 seconds suggests a less efficient model compared to the parameterized probabilistic regression approach in XGBLSS. However, the CPS-Geo model's broader interval ensures that it meets the desired coverage, highlighting a trade-off between precision and reliability.

Conversely, the XGB-OSM model consistently lags in performance for Boston, with a CRPS of 79.5 seconds and a pinball loss at the 50th percentile of 52.4 seconds, demonstrating that it is the least effective model among those evaluated. This relative under performance is indicative of its limitations in capturing the nuances of the urban context compared to the other models.

In Seattle, the 3-ring model again outperforms the City model, albeit not as markedly as in Boston, with a CRPS of 71.4 seconds for the 3-ring versus 74.4 seconds for the City model. This difference aligns with the patterns observed in Figure~\ref{fig:cluster_van_times}, where Boston exhibits a higher variability in service time across urban regions compared to Seattle. Interestingly, all predictive models in Seattle surpass the 3-ring model in CRPS, with the CPS model also outperforming in the pinball loss at the 95th percentile ($L_{0.95}$). Specifically, the CPS-Geo model achieves the best overall performance with a CRPS of 70.2 seconds and a pinball loss at the 95th percentile of 26.3 seconds, indicating its robustness in probabilistic forecasting.\footnote{We note that in the Seattle context, the XGB-OSM model not only outperforms the XGB-Geo model across all metrics but also demonstrates a better generalization capability. The XGB-OSM model achieves a CRPS of 70.4 seconds, besting that of XGB-Geo. This suggests that the hyperparameters tuned for Boston may not be optimal for Seattle, whereas the parameters of the XGB-OSM model appear to generalize better. We were not able to compare Seattle specific hyperparameters at the time of submission, but will update results accordingly.
}

In Table~\ref{tab:city_bike}, we compare city-specific models for cargo bikes, analyzing their performance across two distinct urban landscapes. Here, the relative difference between the 3-Ring and City models is considerably less pronounced than in the van models, which aligns with the observations made in Section \ref{sec:service_times_comparison}. In this section, we noted that cargo bike service times exhibit minimal variability across the urban landscape, likely due to their ability to park an average of 22 meters from the delivery destination. This reduced variability sometimes results in the City model outperforming the 3-Ring model; for instance, in London, it presents a narrower pinball loss at the 95th percentile with 21.9 seconds versus 22.6 seconds.

\begin{table}[h]
    \centering
    \caption{\textbf{City-Specific Van Model Comparison}. The baseline models are separated from the predictive models by a horizontal line. The best value of the baseline models in each metric is underlined, with the best value for the predictive models being bolded.}
    \label{tab:city_bike}
    \begin{tabular}{lllllll}
    \toprule
     &  & \multirow[*]{2}{*}{CRPS (s)} & \multicolumn{2}{c}{Mean Interval} & \multicolumn{2}{c}{Pinball} \\
     City & Model &  & Coverage (\%) & Width (s) & $P_{50}$ (s) & $P_{95}$ (s) \\
    \midrule
    \multirow[*]{5}{*}{London} & 3-Ring & \underline{$77.4\pm2.7$} & $86.6\pm1.8$ & \underline{$412.1\pm10.1$} & \underline{$52.3\pm1.7$} & $22.6\pm1.3$ \\
     & City & $77.7\pm2.3$ & \underline{$90.8\pm0.9$} & $435.0\pm0.0$ & $53.3\pm1.4$ & \underline{$21.9\pm1.6$} \\
     \cline{2-7}
     & CPS-Geo & $\mathbf{77.4\pm2.7}$ & $\mathbf{91.1\pm1.3}$ & $455.9\pm18.5$ & $\mathbf{53.1\pm1.6}$ & $\mathbf{22.6\pm2.1}$ \\
     & XGB-Geo & $82.1\pm3.0$ & $66.5\pm0.9$ & $\mathbf{271.5\pm7.5}$ & $53.9\pm1.8$ & $28.8\pm2.4$ \\
     & XGB-OSM & $78.3\pm2.9$ & $83.4\pm1.4$ & $384.3\pm11.1$ & $53.3\pm1.8$ & $23.7\pm1.8$ \\
    \cline{1-7}
    \multirow[*]{5}{*}{Brussels} & 3-Ring & $67.7\pm2.2$ & $87.5\pm1.0$ & \underline{$371.2\pm2.9$} & $46.7\pm1.5$ & $16.9\pm0.6$ \\
     & City & \underline{$67.1\pm2.1$} & \underline{$91.1\pm1.3$} & $393.0\pm0.0$ & \underline{$46.4\pm1.4$} & \underline{$16.7\pm0.6$} \\
     \cline{2-7}
     & CPS-Geo & $\mathbf{67.8\pm1.6}$ & $\mathbf{90.2\pm2.9}$ & $395.7\pm13.4$ & $46.9\pm1.1$ & $\mathbf{17.2\pm0.5}$ \\
     & XGB-Geo & $70.3\pm1.6$ & $66.1\pm3.0$ & $\mathbf{246.1\pm12.7}$ & $47.5\pm1.3$ & $21.2\pm1.4$ \\
     & XGB-OSM & $68.2\pm1.9$ & $82.8\pm1.4$ & $348.3\pm6.0$ & $\mathbf{46.8\pm1.2}$ & $18.1\pm0.8$ \\
    
    \bottomrule
    \end{tabular}

\end{table}

In London, the CPS-Geo model emerges as the best predictive model with a CRPS of 77.4 seconds, followed by the XGB-OSM model. Interestingly, there are instances where the City average model outperforms the contextual models, such as the City model's mean interval coverage of 90.8\%, indicating that urban context may offer limited predictive value for cargo bike deliveries. This suggests that the inherent advantages of cargo bikes in navigating urban terrain might decresase the impact of using urban context to predict service times.

For Brussels, the pattern observed is similar. The 3-Ring model and the City model are closely matched, with the City model showing a slightly better CRPS of 67.1 seconds compared to the 3-Ring model's 67.7 seconds. The City model also achieves marginally better pinball loss at the 95th percentile with 16.7 seconds versus 16.9 seconds. The CPS model, while not outperforming the City model in CRPS or pinball loss, does provide the best mean interval coverage at 90.2\%, suggesting its effectiveness in capturing the variability of service times. 

These findings indicate that for cargo bikes, the ability to bypass many of the challenges faced by larger delivery vehicles within urban centers, such as parking restrictions and traffic congestion, leads to a more uniform service time. This uniformity reduces the predictive utility of models that heavily rely on urban context, suggesting that simpler models may sometimes suffice. The results also highlight the need for further research into the unique operational characteristics of cargo bikes to better understand and optimize their use in urban delivery systems.

\subsubsection{Cross City Prediction}

The overarching goal of our predictive models is to achieve generalizability across various geographies. In this section we try to understand what the best modelling strategy is to predict service times in a city under different data availability scenarios (i.e. little or no data). We thus evaluate the performance of the van model in three different training scenarios:

\begin{itemize}
    \item \textbf{City-Specific:} The model was exclusively trained on data from Boston and then tested on a separate test set within Boston.
    \item \textbf{Transfer:} The model was trained on delivery data from all U.S. cities except Boston (including Seattle, Austin, and Chicago) and subsequently applied to the test set in Boston.
    \item \textbf{Full:} The model's training encompassed all U.S. cities plus the training set from Boston, and it was then tested on the Boston test set.
\end{itemize}

\begin{table}[h]
        \caption{\textbf{Comparing the performance of different training schemes of the Boston van model}. \textit{Full} represents training on all U.S. cities, including Boston. \textit{City-Specific} was trained only on Boston. \textit{Transfer} was trained on all U.S. cities besides Boston. The best score for each metric is bolded.}
    \label{tab:cross_city}
    \begin{tabular}{lllllll}
    \toprule
     &  & \multirow[*]{2}{*}{CRPS (s)} & \multicolumn{2}{c}{Mean Interval} & \multicolumn{2}{c}{Pinball} \\
     Scheme & Model &  & Coverage (\%) & Width (s) & $P_{50}$ (s) & $P_{95}$ (s) \\
    \midrule
    \multirow[*]{2}{*}{City} & CPS-Geo & $80.5\pm1.8$ & $\mathbf{90.1\pm1.3}$ & $437.0\pm27.4$ & $51.7\pm1.2$ & $29.5\pm0.9$ \\
 & XGB-Geo & $80.7\pm1.8$ & $86.1\pm1.1$ & $350.2\pm11.2$ & $51.8\pm1.0$ & $30.8\pm0.6$ \\
\cline{1-7}
\multirow[*]{2}{*}{Full} & CPS-Geo & $80.4\pm1.5$ & $87.5\pm0.9$ & $394.9\pm13.3$ & $\mathbf{51.5\pm1.0}$ & $\mathbf{29.5\pm0.7}$ \\
 & XGB-Geo & $\mathbf{80.1\pm2.0}$ & $86.4\pm1.1$ & $342.1\pm10.6$ & $51.5\pm1.2$ & $30.4\pm0.8$ \\
\cline{1-7}
\multirow[*]{2}{*}{Transfer} & CPS-Geo & $82.0\pm2.3$ & $86.3\pm1.0$ & $362.7\pm7.3$ & $52.6\pm1.5$ & $32.2\pm1.2$ \\
 & XGB-Geo & $83.2\pm2.6$ & $81.0\pm1.0$ & $\mathbf{278.3\pm2.6}$ & $52.5\pm1.4$ & $36.7\pm2.0$ \\
\bottomrule
\end{tabular}

\end{table}

In Table~\ref{tab:cross_city}, we present a comparative analysis of the models under these training scenarios. The assessment uses the same metrics as those in Section~\ref{sec:city_model}, as well as a 5-fold cross-validation. 

The Full CPS-Geo model stands out, showing the best pinball loss performance and comparable CRPS to the Full XGB-Geo model, which leads in CRPS. These outcomes indicate that there are indeed learnable geospatial relationships that can enhance the predictive capability of a model beyond what is possible with a single city's data. However, a tendency of the Full model to underpredict the width of the Boston intervals suggests that while these relationships exist, the model may not fully capture the higher variance in service times seen in Boston, as detailed in Table~\ref{table:service_times}, instead having a training set biased towards the other cities, it learns a lower variance.

When examining the Transfer scenario, the models showcase their resilience and capability to generalize. Specifically, the Transfer CPS-Geo model registers a CRPS of $82.0\pm2.3$ seconds, which, while not eclipsing the performance of the Full or City-specific models, is a marked improvement over the City baseline model presented in Table~\ref{tab:city_van}, where the City model reported a CRPS of $85.8\pm9.1$ seconds for Boston. This significant enhancement in predictive accuracy—evident in the approximately 4.4-second reduction in CRPS—underscores the effectiveness of the Transfer model in adapting to a new city, despite being trained on datasets from different urban areas.

 These findings, although preliminary, show the potential of probabilistic modeling techniques like CPS and XGB-LSS, especially when utilizing GeoVex embeddings to capture and generalize the urban context's influence on delivery service times. The results highlight the models' adaptability and suggest that with further refinement, they could be applied effectively across a range of urban environments, paving the way for more universally applicable delivery time prediction models.

\section{Conclusion}

The rapid growth of e-commerce and the increasing demand for last-mile deliveries have led to a significant rise in the use of light goods vehicles (LGVs) in cities. However, LGVs are one of the leading polluters in urban areas, contributing to air pollution, carbon emissions, traffic congestion, parking challenges, and are responsible for a disproportionate number of fatal collisions. In response to these issues, light electric vehicles (LEVs) and cargo bike logistics have been presented as high-impact candidates to replace LGVs in cities. 

Despite the potential benefits of LEVs and cargo bikes, their commercial competitiveness is still poorly understood. To accelerate the shift towards more sustainable urban delivery solutions, it is crucial to develop accurate models of delivery performance that account for the complexity of urban environments.
In this study, we introduced a novel framework to measure the delivery performance of vehicles across diverse urban areas. Our focus was on understanding how various urban contexts influence the performance of these different delivery vehicles. 

We defined a computational framework, leveraging Uber’s H3 index to divide cities into hexagonal cells, to allow us to study the behaviour and efficiency of vehicles at the ”micro-region” level. In this paper, we specifically looked at the service time aspect of deliveries, which includes the time spent on activities such as parking, unloading, and walking to delivery points. Despite its considerable importance in urban delivery scenarios,  this is area that has not received extensive attention in urban logistics research.

We analysed a public dataset of Amazon van deliveries, and introduced two new cargo-bike datasets based in London and Brussels. Our modelling framework leveraged state-of-the-art geospatial embedding to represent urban context. We used this to predict service times across urban micro-regions. We demonstrated our model's ability to generalise to new and unseen urban contexts. This is particularly important given the difficulty of collecting data in the urban logistics industry.

Overall, our findings indicate that factors like building density, road types, and local amenities are crucial in determining the service times of delivery vehicles. In the Amazon dataset, LGVs were significantly impacted by dense city centres. On the other hand, cargo-bikes, benefiting from their smaller size and greater manoeuvrability, were comparably unaffected by these challenges. Indeed, we found consistently small distances to the address for cargo-bikes, pointing at minimal time spent cruising and walking to the door.

However, our study was limited by distinctive operational patterns between our datasets, preventing a direct comparisons between the datasets. Our research faced limitations due to the level of detail in the data available, specifically with the absence of specific details about walking times and distance to the door for vans. Despite the challenges in data granularity, our results support that cargo-bikes may have significant advantages in dense urban areas. 

Overall, the significance of our work lies in defining a framework for assessing vehicle performance in urban deliveries. Future research will aim to develop more comprehensive models that account for a variety of factors affecting urban delivery vehicles beyond just service times, such as loading times, and vehicle-specific navigation and journey times, providing a more accurate view of the efficiency and practicality of different delivery methods in urban settings. We also aim to tackle the operational challenges related to light electric vehicles, investigating how their limited carrying capacity might be offset by optimising urban infrastructure use, like micro-hubs, to facilitate reloading.

\section*{Declarations}
\backmatter

\bmhead{Funding}

This work was supported by Climate Change AI through its Innovation Grant Programme.


\bmhead{Data Availability Statement}

Code and datasets are availbale on the Github repository: https://github.com/green-last-mile/cargo-bike-analysis


\bmhead{Ethics approval}
We analyzed anonymized delivery data, not requiring ethics approval due to its non-identifiable nature, yet complied with GDPR for data protection.

\bmhead{Consent to participate}
Not applicable.

\bmhead{Consent for publication}
We received explicit consent from Pedal Me and Urbike for the publication and sharing of the data used in this study. This consent covers the use of anonymized data for research purposes and its dissemination in academic publications.









\begin{appendices}

\clearpage
\section{Completeness of the OpenStreetMap dataset}\label{appendix:OSM}

Despite the feature-rich nature of the OpenStreetMap (OSM) data, concerns have been raised over its quality, owing in part to the open-source nature, but also to the ever-changing nature of geography.

\textbf{The positional accuracy of OSM data}.
There has been a growing body of research focused on evaluating the positional accuracy of OSM data. One study found that the quality of OSM buildings in Lombardy, Italy was comparable to that of the regional technical authoritative map at a scale of 1:5000, or below the threshold of 3 meters of error~\cite{brovelli2018new}. Similarly, a study in Munich and found that the average difference between the buildings in OSM and a government database was only 4m~\cite{fan2014quality}. In general, research has found that buildings with the highest positional accuracy were primarily located in more populous regions of the city and suggest that the level of validation and correction that OSM data undergoes may have a significant impact on its accuracy, with areas subject to more frequent validation and error correction exhibiting higher levels of accuracy~\cite{helbich2012comparative}. As our work is only interested in OSM objects at an aggregate level, we did not investigate the positional accuracy further, however we too observe that populous regions have more complete data and apparent accuracy.

\textbf{The completeness of OSM data}.
OSM reminds users that \textit{the map is not finished yet}~\footnote{\url{https://wiki.openstreetmap.org/wiki/About_OpenStreetMap}}. Research has found the data to be generally more complete and up-to-date in urban regions compared to rural areas~\cite{hecht2013measuring, helbich2012comparative}. Still, the completeness can vary greatly between cities and regions. For example, Munich had near 100\% coverage of buildings~\cite{fan2014quality} in 2014, whereas the entire region of Lombardy, Italy was missing 57\% percent of buildings when compared to the regions official dataset in 2018~\cite{brovelli2018new}.

\begin{figure}[h]
\centering
\includegraphics[width=0.6\textwidth]{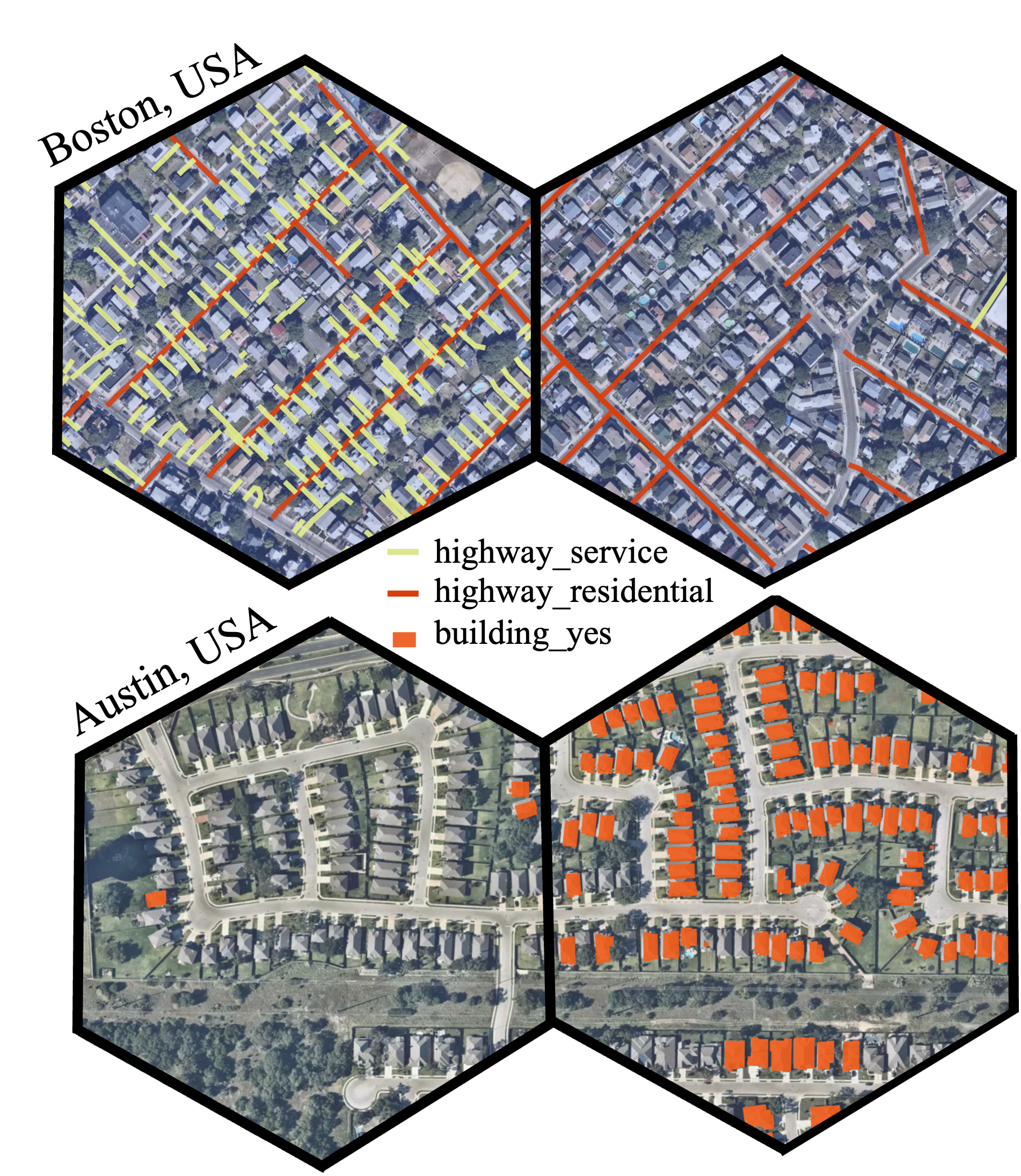}
\caption{Display of neighbor H3 cells in Boston (\texttt{892a3075d77ffff}, \texttt{892a3075d3bffff}) and Austin (\texttt{89489e24d7bffff}, \texttt{89489e248b7ffff}) with visual similarity but disparate OSM tags.}
\label{fig:problem_hexagons}
\end{figure}

We have observed the same variability in completeness. In the Amazon dataset, there are 233 deliveries to a municipality in the Chicago, Illinois suburbs called Berwyn, Illinois. It covers $10.1 km^2$ and has a population density of $5661/km^2$, making it the most population dense municipality in Illinois. The 2020 census reports 21,037 housing units, but OSM only has 8.4\%, or 1775 of Berwyn buildings recorded\cite{chicago_census}. In fact, comparing OSM and satellite imagery for the Chicago area reveals large swaths of missing buildings. Still, Berwyn is not the only example of missing data. In Figure~\ref{fig:problem_hexagons}, we present two pairs of neighbor H3 cells. The top two are located in the greater Boston, Massachusetts area and the bottom in Austin, Texas. The Boston pair highlights how geographically close H3 cells (their centers are 362 meters apart) with similar satellite imagery can have significantly different OSM tagging schemes. \texttt{892a3075d77ffff} has each house's driveway labelled as \texttt{highway:service}. The neighbor, \texttt{892a3075d3bffff}, does not have labelled driveways at all, resulting in 167 \texttt{highway:service} tags in the left cell and the 3 in the right. The bottom cells in Austin again illustrate the problem of missing buildings, with the left cell having 3 building tags and the right cell 110.

\textbf{The semantic meaning of OSM tags}
As explained, OSM relies on their users to categorize geographic entities through the use of \texttt{key=value} pairs (also referred to as tags). They list the most popular tags on their wiki, however there is no formal restriction in place on users - meaning that tags and values can be freely created~\cite{vandecasteele2015improving}. To deal with the open-ended nature, literature suggests only using OSM entries which has a tag present on the OSM wiki page~\footnote{\url{https://wiki.openstreetmap.org/wiki/Map_features}}~\cite{wozniak_hex2vec_2021}. However, even with outlier tags filtered out, they still contain semantic heterogeneity. There are several reasons for the heterogeneity cited in literature, with one of the common causes being that the geographic entity has high ambiguity, \textit{i.e.} \texttt{amenity=gym} vs. \texttt{leisure=fitness}. Another owes to the fact that there are geographic differences in the way that tags are used, depending on localized naming conventions or the interpretation of the scale of the object itself, \text{i.e.} \texttt{water=pond} vs. \texttt{water=lake}~\cite{vandecasteele2015improving}.

\begin{figure}[h]
\centering
\includegraphics[width=0.8\textwidth]{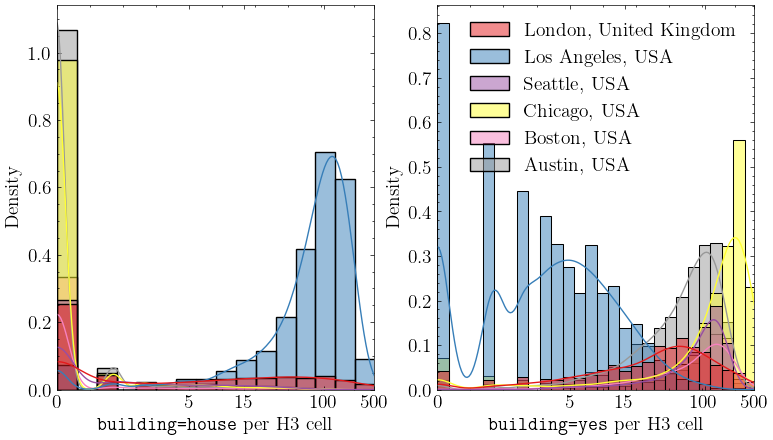}
\caption{Distribution of the count of \texttt{building\_house} vs \texttt{building\_yes} tags in the 6 cities of interest. Los Angeles, USA has a high frequency of 0 \texttt{building\_yes} tags per hexagon.}
\label{fig:building_yes_diff}
\end{figure}

We observe semantic differences in our cities of interest as well. Figure~\ref{fig:building_yes_diff} highlights the tagging differences for buildings in our study cities. OSM contributors for Los Angeles, California have labelled single family homes as \texttt{building=house}, which is apparent by the high frequency of H3 cells with non-zero \texttt{building=house} counts. Single family homes in the other cities are largely labelled with the catch-all building tag, \texttt{building=yes}. The Los Angeles tagging scheme provides more context and enriches the semantic meaning locally~\cite{vandecasteele2015improving}, however when expanding the scope beyond Los Angeles, the skewed distribution of \texttt{building:house} counts makes Los Angeles cells outliers.

\textbf{OSM data problems highlighted in existing works} 
Although the OSM data is constantly under improvement, the completeness and quality of the annotations in different regions are affected by the number and mapping skills of the volunteers \cite{mooney2012annotation}. As reported in \cite{haklay2010good}, the spatial coverage of OSM is heterogeneous in different geographical regions — i.e., urban areas are more regularly updated than rural areas. In road networks, missing roads are reported in \cite{funke2015automatic} and inaccurate road tags are reported in \cite{jilani2014automated}. The positional accuracy of building footprints in OSM sometimes requires corrections \cite{xu2017quality}. Several works in the literature have studied methods to assess the quality of OSM data by quantifying: data completeness \cite{koukoletsos2012assessing}, positional accuracy \cite{fan2014quality}, semantic tag acccuracy \cite{girres2010quality}, and topological consistency \cite{neis2011street}. Some works focus on meta analysis of OSM, like the analysis of the contributors’ activities \cite{neis2014recent, arsanjani2015exploration} and the quality assessment of the OSM data \cite{senaratne2017review, jilani2019traditional}.

Despite its issues, OSM remains the largest and most content-rich open-source geographic dataset, providing researchers with valuable insights into global geography. Recently, researchers have utilized strategies from natural language processing and representation learning to address issues of completeness and accuracy with OSM tags.
\clearpage
\section{City descriptions through models of urban micro-regions and Large Language Models}\label{Appendix:city-descriptions}

In this section, we provide high level descriptions of the make up of the different cities studied across our data sets. Inspired by the work of \cite{yap2023global}, we use chatGPT (GPT-4) to assist in generating high level descriptors of the different OSM tags present in each cluster~\cite{van2023chatgpt}. Using a regex mapping function, we translate each OSM tag into one of the nine following "super-tags": Built Environment, Transportation, Natural Elements, Amenities, Leisure \& Recreation, Barriers \& Boundaries, Utilities \& Services, Commerce \& Industry, and Historical \& Cultural. Based on the super-tag make up of each cluster, we use chatGPT to label and generate descriptions of each urban region (cluster) across all six cities. We also collect population data from Meta’s high-resolution population density maps, which provide spatially detailed population data in 30-m spatial resolution for 200 countries, through the \emph{urbanity} python library~\cite{yap2023global}. 

\subsubsection*{Boston, USA}

\begin{table}[ht!]
\centering
\small 
\caption{Urban Cluster Characteristics in Boston, USA}
\label{tab:boston_clusters}
\begin{tabularx}{\textwidth}{@{}l *{9}{X}@{}}
\toprule
\textbf{Cluster} & \textbf{Hex} & \textbf{Pop.} & \textbf{Amen.} & \textbf{Built Env.} & \textbf{Comm. \& Ind.} & \textbf{Hist. \& Cult.} & \textbf{Leis. \& Rec.} & \textbf{Nat. Elem.} & \textbf{Transp.} \\ \midrule
Downt. Core & 98 & 819.48 & 34.26 & 105.44 & 4.34 & 1.56 & 7.93 & 0.72 & 180.11 \\
Out. Urb. Res. & 159 & 752.18 & 13.18 & 75.89 & 2.02 & 0.54 & 5.45 & 0.69 & 107.06 \\ 
Mix. Use & 87 & 554.61 & 12.34 & 91.87 & 2.23 & 0.71 & 4.84 & 1.17 & 99.31 \\
Suburb Res. & 795 & 472.96 & 3.30 & 111.05 & 0.45 & 0.22 & 2.37 & 0.61 & 50.34 \\

\bottomrule
\end{tabularx}
\end{table}

In the context of Boston, our within-city clustering delineates four distinct urban clusters. The "Suburban Residential" cluster, with a significant hex count of 795, is characterized by a moderate population and a focus on residential amenities, reflective of a suburban environment. In contrast, the "Downtown Core" cluster, though comprising only 98 hexes, is densely populated with a high concentration of amenities and commerce, epitomizing the urban center's vibrancy. The "Mixed-Use Area" cluster, encompassing 87 hexes, balances residential and commercial aspects, indicating areas of mixed functionality. It is marked by moderate population and commerce presence. Finally, the "Outer Urban Residential" cluster, with 159 hexes, presents higher population density with a predominance of residential features, suggesting peripheral urban living spaces.

\subsubsection*{Seattle, USA}

\begin{table}[ht!]
\centering
\small 
\caption{Urban Cluster Characteristics in Seattle, USA}
\label{tab:seattle_clusters}
\begin{tabularx}{\textwidth}{@{}l *{9}{X}@{}}
\toprule
\textbf{Cluster} & \textbf{Hex} & \textbf{Pop.} & \textbf{Amen.} & \textbf{Built Env.} & \textbf{Comm. \& Ind.} & \textbf{Hist. \& Cult.} & \textbf{Leis. \& Rec.} & \textbf{Nat. Elem.} & \textbf{Transp.} \\ \midrule
Downtown Core & 497 & 349.96 & 15.12 & 104.01 & 3.30 & 0.30 & 3.34 & 0.58 & 149.18 \\
Com. \& Ind. Hub & 473 & 339.74 & 10.59 & 136.51 & 3.32 & 0.23 & 2.84 & 0.57 & 102.14 \\
Urb. Residential & 325 & 232.07 & 2.54 & 94.74 & 0.81 & 0.18 & 2.09 & 0.67 & 39.31 \\ 
Suburban Residential & 981 & 204.57 & 3.72 & 87.99 & 1.38 & 0.13 & 1.83 & 0.52 & 59.60 \\
\bottomrule
\end{tabularx}
\end{table}

In Seattle, our within-city clustering analysis reveals four distinct clusters, each representing a unique urban fabric. The "Downtown Core" cluster, with 497 hexes, is characterized by a high population density, ample amenities, and significant built environment, showcasing the bustling nature of the city center. This cluster also has the highest transportation index, indicating its role as a primary transit hub. In contrast, the "Suburban Residential" cluster, the largest with 981 hexes, exhibits lower population density and amenities, typical of suburban areas. This cluster is defined by its residential nature, with less emphasis on commerce and industry. The "Commercial and Industrial Hub" cluster, comprising 473 hexes, stands out with the highest built environment index, reflecting its focus on commerce and industry. This cluster, with a moderate population density and ample amenities, represents areas primarily dedicated to business and commercial activities. Lastly, the "Urban Residential" cluster, with 325 hexes, displays characteristics of urban residential areas. It has a moderate population density and a lower emphasis on amenities and commerce, indicative of residential neighborhoods within the urban periphery.

\subsubsection*{Austin, USA}

\begin{table}[ht!]
\centering
\small 
\caption{Urban Cluster Characteristics in Austin, USA}
\label{tab:austin_clusters}
\begin{tabularx}{\textwidth}{@{}l *{9}{X}@{}}
\toprule
\textbf{Cluster} & \textbf{Hex} & \textbf{Pop.} & \textbf{Amen.} & \textbf{Built Env.} & \textbf{Comm. \& Ind.} & \textbf{Hist. \& Cult.} & \textbf{Leis. \& Rec.} & \textbf{Nat. Elem.} & \textbf{Transp.} \\ \midrule
Urban Core & 620 & 270.02 & 6.39 & 116.91 & 1.47 & 0.22 & 2.74 & 1.21 & 47.40 \\
Developed Residential & 809 & 255.25 & 3.68 & 67.97 & 1.81 & 0.07 & 1.89 & 1.11 & 39.51 \\
Emerging Urban & 1804 & 219.41 & 1.36 & 78.81 & 0.56 & 0.07 & 1.62 & 1.42 & 29.82 \\
Rural/Suburban & 1889 & 103.40 & 0.62 & 39.77 & 0.20 & 0.02 & 1.14 & 1.20 & 16.03 \\

\bottomrule
\end{tabularx}
\end{table}

In Austin, our within-city clustering uncovers four distinct clusters, each reflecting a unique aspect of the city's urban landscape. The "Rural/Suburban" cluster, with the largest hex count of 1889, is characterized by its low population density and minimal amenities, indicative of rural or suburban areas with sparse development. The "Emerging Urban" cluster, with 1804 hexes, shows signs of urban development, evidenced by a higher population density and increased amenities. This cluster likely represents areas in transition from suburban to more urban settings. The "Urban Core" cluster, though smaller with 620 hexes, has the highest population density and the most developed built environment. This cluster is the heart of urban activity in Austin, with a significant presence of amenities, commerce, and industry. Finally, the "Developed Residential" cluster, comprising 809 hexes, reflects established residential areas within the city. It combines a higher population density with moderate amenities and built environment, suggesting well-developed residential neighborhoods.

\subsubsection*{Chicago, USA}

\begin{table}[ht!]
\centering
\small 
\caption{Urban Cluster Characteristics in Chicago, USA}
\label{tab:chicago_clusters}
\begin{tabularx}{\textwidth}{@{}l *{9}{X}@{}}
\toprule
\textbf{Cluster} & \textbf{Hex} & \textbf{Pop.} & \textbf{Amen.} & \textbf{Built Env.} & \textbf{Comm. \& Ind.} & \textbf{Hist. \& Cult.} & \textbf{Leis. \& Rec.} & \textbf{Nat. Elem.} & \textbf{Transp.} \\ \midrule
Downtown Core & 600 & 687.54 & 12.74 & 102.42 & 4.12 & 0.57 & 5.63 & 1.02 & 115.83 \\
Industrial Hub & 1132 & 349.81 & 2.45 & 165.81 & 1.76 & 0.26 & 1.72 & 0.55 & 46.96 \\
Urban Residential & 1622 & 578.55 & 5.17 & 195.38 & 2.78 & 0.59 & 2.06 & 1.25 & 99.78 \\
Suburban Residential & 1448 & 493.63 & 2.95 & 233.37 & 1.95 & 0.36 & 1.71 & 0.36 & 45.43 \\ 
\bottomrule
\end{tabularx}
\end{table}

In Chicago, the within-city clustering reveals a diverse urban composition across four clusters. The "Downtown Core" cluster, with 600 hexes, is characterized by the highest population density and amenities, symbolizing the city's bustling center. This cluster also exhibits substantial built environment and transportation infrastructure, aligning with the typical features of a city's heart. The "Industrial Hub" cluster, encompassing 1132 hexes, shows a lower population density but a significant built environment, suggesting areas with industrial and commercial emphasis. The moderate presence of amenities and transportation facilities supports this characterization. In contrast, the "Urban Residential" cluster, the largest with 1622 hexes, balances residential characteristics with urban development. This cluster has a high built environment index and a moderate population density, indicative of dense residential areas within the urban fabric. Lastly, the "Suburban Residential" cluster, comprising 1448 hexes, is marked by a relatively high built environment index with lower amenities and commerce, reflecting the features of suburban residential areas.

\subsubsection*{Brussels, Belgium}
\begin{table}[ht!]
\centering
\small 
\caption{Urban Cluster Characteristics in Brussels, Belgium}
\label{tab:brussels_clusters}
\begin{tabularx}{\textwidth}{@{}l *{9}{X}@{}}
\toprule
\textbf{Cluster} & \textbf{Hex} & \textbf{Pop.} & \textbf{Amen.} & \textbf{Built Env.} & \textbf{Comm. \& Ind.} & \textbf{Hist. \& Cult.} & \textbf{Leis. \& Rec.} & \textbf{Nat. Elem.} & \textbf{Transp.} \\ \midrule
Urban Core & 416 & 580.66 & 15.31 & 119.99 & 2.81 & 0.27 & 5.83 & 2.55 & 120.83 \\
Cultural Hub & 427 & 1254.58 & 31.56 & 268.47 & 8.93 & 1.13 & 6.88 & 1.49 & 136.34 \\
Suburban Area & 468 & 351.36 & 4.60 & 98.57 & 1.04 & 0.17 & 4.49 & 2.70 & 43.43 \\
Mixed-Use Area & 427 & 669.44 & 9.61 & 168.59 & 2.18 & 0.31 & 5.60 & 1.63 & 77.08 \\ 
\bottomrule
\end{tabularx}
\end{table}

In Brussels, our clustering analysis delineates four unique urban clusters. The "Urban Core" cluster, consisting of 416 hexes, is marked by a moderate population density and high levels of amenities and transportation, indicating the central, bustling part of the city. The "Cultural Hub" cluster, with 427 hexes, stands out with the highest population density and the most significant presence of amenities, historical and cultural elements, and built environment. This cluster likely represents areas with a rich cultural heritage and vibrant city life. The "Suburban Area" cluster, comprising 468 hexes, features lower population density and amenities. This cluster is characterized by more residential and natural elements, typical of suburban regions on the outskirts of the city. Lastly, the "Mixed-Use Area" cluster, also with 427 hexes, balances residential and commercial characteristics. It shows a higher built environment index and a moderate population, suggesting areas that combine living and commercial spaces within the urban landscape.

\subsubsection*{London, UK}

\begin{table}[ht!]
\centering
\small 
\caption{Urban Cluster Characteristics in London, UK}
\label{tab:london_clusters}
\begin{tabularx}{\textwidth}{@{}l *{9}{X}@{}}
\toprule
\textbf{Cluster} & \textbf{Hex} & \textbf{Pop.} & \textbf{Amen.} & \textbf{Built Env.} & \textbf{Comm. \& Ind.} & \textbf{Hist. \& Cult.} & \textbf{Leis. \& Rec.} & \textbf{Nat. Elem.} & \textbf{Transp.} \\ \midrule
Urban Center & 1202 & 621.96 & 10.18 & 126.14 & 3.81 & 0.41 & 8.26 & 1.45 & 63.03 \\ 
Cultural Core & 1331 & 1143.25 & 33.60 & 133.79 & 11.86 & 1.31 & 10.30 & 1.49 & 108.19 \\
Urban Residential & 1671 & 1036.78 & 14.12 & 133.42 & 6.10 & 0.66 & 6.61 & 1.13 & 67.46 \\
Suburban & 2415 & 622.26 & 5.14 & 56.46 & 2.03 & 0.31 & 2.72 & 1.61 & 43.10 \\
\bottomrule
\end{tabularx}
\end{table}

In London, our clustering analysis reveals a diverse and complex urban landscape. The "Cultural Core" cluster, with 1331 hexes, is characterized by the highest population density and a significant presence of amenities, built environment, and historical and cultural elements. This cluster likely represents the city's central areas rich in cultural and historical significance. The "Urban Residential" cluster, the largest with 1671 hexes, exhibits high population density with a substantial built environment, indicative of densely populated residential areas within the urban fabric. The "Urban Center" cluster, consisting of 1202 hexes, strikes a balance between residential and commercial characteristics. It shows a high built environment index and a moderate population density, suggesting areas with a mix of living and commercial spaces. Finally, the "Suburban" cluster, encompassing 2415 hexes, displays lower population density and built environment indices, characteristic of suburban areas on the city's outskirts. This cluster has a more residential and natural focus, with lesser commercial and industrial activity.




\end{appendices}


\bibliography{sn-bibliography}

\end{document}